\documentclass[reprint,prb,showpacs,preprintnumbers,amsmath,amssymb,twocolumns,footinbib,citeautoscript,superscriptaddress]{revtex4-1}

\usepackage{dcolumn}     
\usepackage{bm}          
\usepackage{graphicx}
\usepackage{epsfig}
\usepackage{epstopdf}
\usepackage{xcolor}

\begin{document}

\newcommand{\uam}{Departamento de F\'{i}sica Te\'{o}rica de la Materia Condensada, Universidad Aut\'{o}noma de Madrid, E-28049 Madrid, Spain}
\newcommand{\ifimac}{Condensed Matter Physics Center (IFIMAC), Universidad Aut\'{o}noma de Madrid, E-28049 Madrid, Spain}

\title{Platinum atomic contacts: from tunneling to contact}

\author{Linda A. Zotti}\affiliation{\uam}
\author{Rub\'{e}n P\'{e}rez}\affiliation{\uam} \affiliation{\ifimac}


\date{\today}

\begin{abstract}
We present a theoretical study of the electronic transport through Pt nanocontacts. We show that
the analysis of the tunnelling regime requires a very careful treatment of the technical details.
For instance, an insufficient size of the system can cause unphysical charge oscillations to arise
 along the transport direction; moreover, the use of an inappropriate basis set can deviate
 the distance dependence of the conductance from the expected exponential trend.
While the conductance decay can be either corrected by employing ghost atoms or a large-cutoff-radius
basis set, the same does not apply to the corrugation, for which only the second option is recommended.
 Interestingly, these details were not found to have a remarkable impact in the contact regime.
These findings are important for theoretical studies of distance-dependent phenomena in
 scanning-probe and break-junction experiments.
\end{abstract}


\maketitle

\section{Introduction}
Over the past decade, metal nanocontacts have  attracted remarkable interest as they make it
 possible to study the electronic, mechanical and thermal properties of metal atoms in
 low-coordination conditions,
\cite{sattler2010handbook,evangeli2015,cuevas2010,jelinek2005,jelinek2008,evangeli2015,Muller2016}
 besides offering the possibility of being operated as two-terminal switches.\cite{schirm2013}
Furthermore, a thorough understanding of their electron transport properties is crucial
in molecular electronics, where the total transmission through metal-molecule-metal junctions 
often includes contribution from metal-to-metal current. \cite{cuevas2010}
Investigating metal contacts at the nanoscale has been possible especially  thanks
 to the refinement of scanning tunnelling microscopy (STM) and
 mechanically-controlled-break-junction (MCBJ) techniques.
In particular, studies on Pt contacts have revealed interesting effects. For example,
it was observed that, upon elongation, the conductance oscillates and the number
of scattering channels changes~\cite{,fernandez2005,garcia2005,vardimon2014}.
 In addition, it was suggested that Pt
is a good candidate as electrode material in molecular junctions as it makes
rectification independent of stretching distances~\cite{parashar2014}.
The trapping of molecules in Pt-Pt junctions, such as CO, has also been studied quite extensively
 and is still the object of interesting analyses~\cite{makk2012pulling}.
Nevertheless, the amount of theoretical studies on electron transport through
 Pt nanocontacts
\cite{garcia2004,strange2006,strange2008,cuevas2003,garcia2005,zhang2010,nielsen2003,nielsen2002,
wu2014,thygesen2005,pauly2006,smogunov2008,Ono2009}
 still remains limited compared, for instance, to gold. The latter is
 more widely used in experiments due to its noble character and, consequently, has often been
employed as staple electrode material in theoretical simulations.
 Furthermore, the Au electronic structure is particularly convenient for transport calculations, as we proceed to explain.
The most common tool to compute electron transmission is the combined
 DFT(density functional theory)+NEGF(Non-Equilibrium Green's Functions) method,
\cite{Pauly2008a,alacant,brandbyge2002,rocha2006spin,ozaki2010,Xue2001,ferrer2014,GarciaLekue2015292} which is
by now implemented in several codes. 
Many of them include
 self consistent cycles in which convergence must be reached for input and output charge density.
Such a task is easier on Au structures than on their Pt
 analogues, due to the fact that the Au valence band consists of the 6$s$ orbital mainly,
 while that of Pt consists of both the 6$s$ and 5$d$ bands.
The limitations and inaccuracies of DFT are well known, as is the huge effort
 which is currently being made to try to overcome them by applying corrections or using
 alternative theoretical
 techniques.\cite{Quek2007} Nevertheless, the DFT+NEGF method is still widely used,
 especially for large systems.
 In this piece of work, we will show theoretical results based on benchmark
electron transport calculations on Pt nanocontacts. We will discuss how to perform
 them in an  accurate manner; furthermore,
  we will examine whether strategic approximations can be made to make these calculations
less cumbersome without jeopardizing the quality of the final result.
 In particular, we will focus on the
 transition from tunnelling to
 contact regime, which has not been studied to the same extent as contacted systems
\cite{limot2005,garcia2010} but which is extremely important when studying
 distance-dependent phenomena in STM or MCBJ experiments. For instance, it was
shown that a change in conductance due to inelastic effects makes it possible to
 characterize the crossover from tunneling to contact;  \cite{Frederiksen2007}
it was also observed, in spin-polarized system, that the transmission probabilities
of the eigenchannels present a nonmonotonic behavior as a function of
the tip-adatom separation. \cite{Polok2011}

\section{Methodology}
We performed electron transport calculations by means of the NEGF  extension of the
 OpenMX code~\cite{ozaki2010}, which is by now widely used.~\cite{sharma2013,okuno2012,ansari2012,lan2015electronic,hashmi2015spin,jippo2015electronic}
The model implemented in this software makes use of periodic boundary conditions. 
The basis sets employed consist of linear combinations of atomic orbitals generated by a confinement
scheme which yields wave functions with zero value beyond a chosen cutoff radius.~\cite{ozaki2003,ozaki2004} 
For all elements, basis sets which have  already been conveniently predefined and optimized are provided
by the software distribution. Following the OpenMX notation, we will define
the basis sets by combinations of $l{x}$ sets, $l$ being the orbital and $x$ being the number of
primitive orbitals used for the construction.

To describe the Pt electronic structure we either used the minimal basis set s1p1d1
(containing 5${p}$, 5${d}$ and 6${s}$ orbitals) or the s2p2d1 basis set
with a cutoff radius of 7 Bohr, if not stated otherwise. Notice that, by constructing the $p$
orbital with two primitive functions,  we have taken polarization into account,
since 6${p}$ orbital is the lowest unoccupied orbital.
In all cases, the exchange-correlation functional GGA (PBE13)~\cite{perdew1997} was employed
and the integration over $k$ space was numerically performed by using a Monkhorst-Pack mesh
\cite{Monkhorst}, where the number of $k$ points was varied depending on the specific case.
For the numerical integrations and the solution of the Poisson equation an energy cutoff of
 200 Ry was used. All calculations 
were converged until the deviation of the energy eigenvalue reached a value lower than 10$^{-6}$ H.

\begin{figure}[t]
\begin{center}
\includegraphics[width=8.2cm]{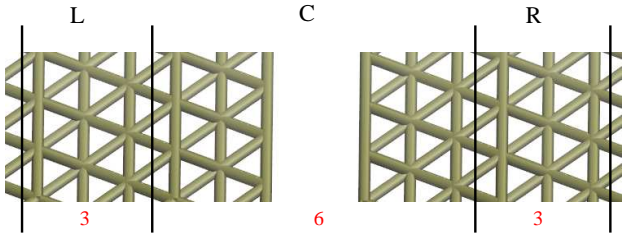}
\caption{ Geometry used to analyze the Mulliken charges in the $3-6-3$ system of
 Fig.~\ref{fig:mulpo}. The vertical black lines separate the central (C), left (L) and right
(R) region. } \label{fig1-slab}
\end{center}
\end{figure}

\section{Results and Discussion} \label{sec-results}
In OpenMX, as in most DFT+NEGF based codes, the whole structure is thought of as divided into
three regions, namely left (L), central (C) and right (R) (Fig.\ref{fig1-slab}).
The electronic transmission is evaluated by the Landauer formula.
Defining k as the Bloch wave vector and $\sigma$ as the spin index, the spin resolved
 transmission is given by
\begin{equation}
T_{\sigma}(E)=\frac{1}{V_{c}} \int_{BZ}{dk^{3}T_{\sigma}^{k}(E)}
\label{T-eq1}
\end{equation}
where $T_{\sigma}^{k}$ is the $k$ resolved transmission defined by
\begin{equation}
T_{\sigma}^{(k)}(E)=Tr[\Gamma_{\sigma,L}^{(k)}(E)G^{(k)}_{\sigma,C}(E+i\epsilon)\Gamma_{\sigma,R}^{(k)}(E)G^{(k)}_{\sigma,C}(E-i\epsilon)]
\label{T-eq2}
\end{equation}
$\Gamma_{\sigma,L}$ and $\Gamma_{\sigma,R}$ are the coupling matrices to the left and right lead.
  The Green's function of the central C region can be written by
\begin{equation}
G^{(k)}_{\sigma,C}(Z)=[ZS^{(k)}_{C}-H^{(k)}_{\sigma,C}-\Sigma^{(k)}_{\sigma,L}(Z)-\Sigma^{(k)}_{\sigma,R}(Z)]^{-1}
\label{G-eq3}
\end{equation}
where the self-energies $\Sigma^{(k)}_{\sigma,L}$ and $\Sigma^{(k)}_{\sigma,R}$ are defined by
\begin{equation}
\Sigma^{k}_{\sigma,L}(Z)=(ZS^{(k)}_{CL1}-H^{(k)}_{\sigma,CL1})G^{(k)}_{\sigma,L(Z)}(ZS_{L1C}^{(k)}-H^{(k)}_{\sigma,L1C})
\label{S-eq4}
\end{equation}
More details about these equations can be found in Ref.~\onlinecite{ozaki2010}.
$G^{(k)}_{\sigma,L(Z)}$ is the surface Green function, which is calculated by a preliminary
standard DFT calculation. This is performed on a bulk structure which has the same chemical composition of the electrode
and which is repeated periodically in all three space directions. 

The first technical issue that has to be addressed is how large each region has to be along the transport direction in order to get a converged value of the conductance. This question is common to other NEGF codes such as 
tranSIESTA \cite{brandbyge2002} and ATK~\cite{ATK}, where convergence has been studied as a function of the size of the central area.
Due to differences in the generation of the basis sets between these codes and OpenMX, such as in the definition of the core potential in the confinement scheme~\cite{junquera2001,ozaki2003,ozaki2004}, it is by no means obvious that the very same criteria should apply to our systems. 
Therefore,  we have applied systematic changes to the size of both the leads and the central part of a Pt$-$Pt junction and checked 
how these changes affect to the electronic structure, focusing in particular to the presence of unphysical charge oscillations 
that can arise close to the leads and which can ultimately affect the computed conductance.

\begin{figure}[t]
\begin{center}
\includegraphics[width=8.2cm]{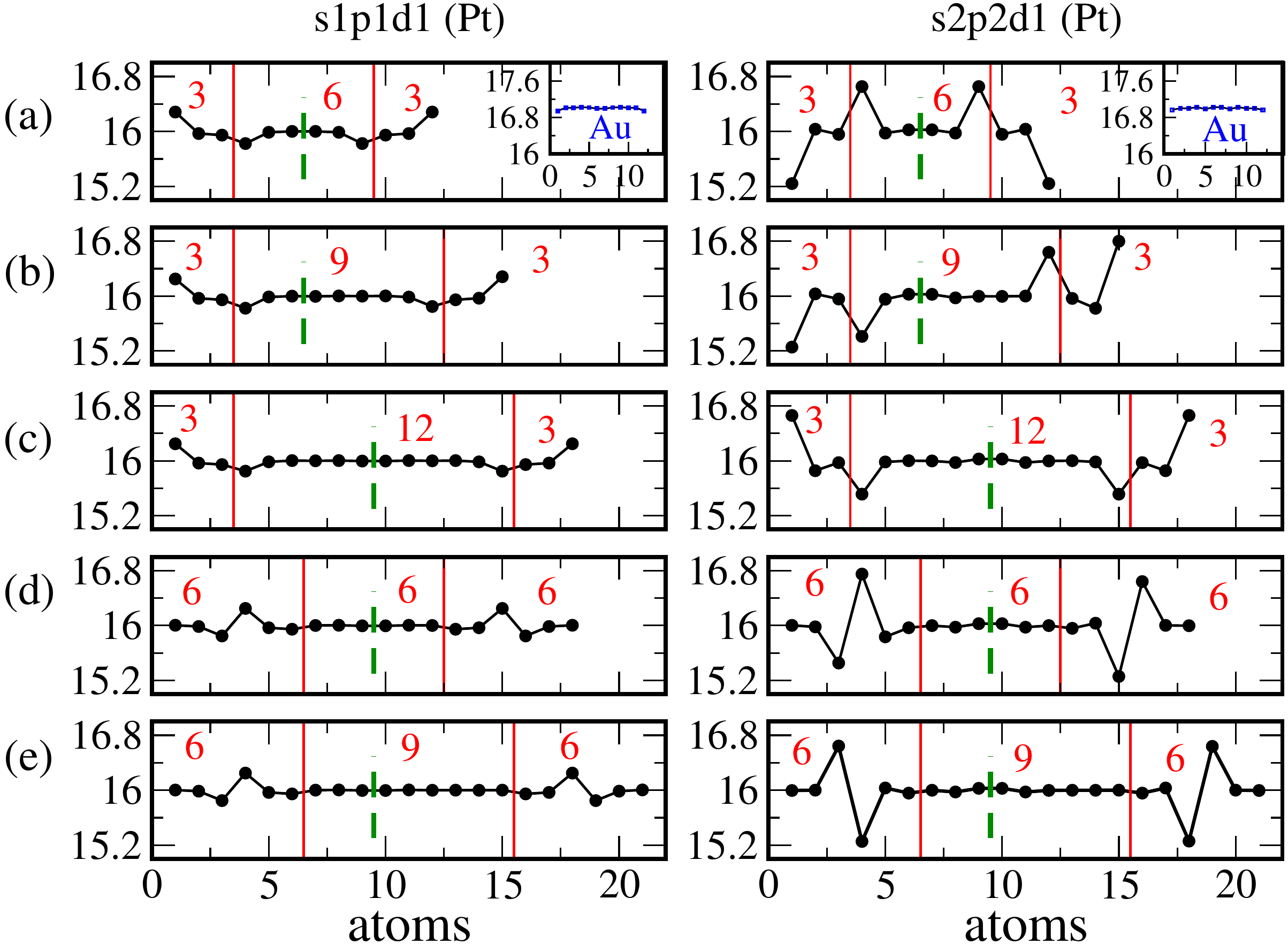}
\caption{(Color online) Mulliken population with (Left)
s1p1d1 and (Right) s2p2d1 basis set
in systems $l-c-r$, where $l$($r$) and $c$ indicate the number of layers in each lead and in the central part, respectively. The following combinations are shown: 3-6-3 (a), 3-9-3 (b),
3-12-3 (c), 6-6-6 (d), 6-9-6 (e). The vertical
dashed green light indicates the 5 {\AA} separation.} \label{fig:mulpo}
\end{center}
\end{figure}

As a first example, we shall analyze the case of the transport through two Pt slabs separated by 5~\AA\  distance along the $\left\{111\right\}$ direction. A periodically repeated 1$\times$1 unit cell was considered, with an 8$\times$8$\times$1 $k$-point grid (Fig.\ref{fig1-slab}).
Notice that the coordination of all atoms is large enough for magnetism to be absent~\cite{fernandez2005}.
Fig.~\ref{fig:mulpo} shows the Mulliken population on each atom along the transport direction for different combinations $l-c-r$ for the number of
layers in the central region ($C$) and in the left ($L$) and right leads ($R$)
 (red numbers in panels a-e). More precisely, $l$ and
$r$ indicate the number of layers in the bulk structure, employed for the preliminary
left and right electrode calculation, which are repeated infinitely along the transport direction.
 In Fig.~\ref{fig:mulpo}, the three areas are separated by
vertical red lines, while the dashed green line indicates the vacuum separation.
The Mulliken charges are shown for both s1p1d1 (left) and s2p2d1 (right) basis set.
The avarage value of 16 (higher than the 10-electron valence of Pt) is due to having
included the electrons of the 5$p$ orbital.
It can be observed that strong charge oscillations (up to 0.8 $e$) appear in proximity of
the leads, especially with the s2p2d1 basis set.This suggests that, while an improvement of 
the basis set is naively expected to yield better-quality results,
in this case it can actually worsen other details such as the
charge population (it is worth noting that improving the quality of the basis set was found to
give qualitatively wrong conclusions also in previous NEGF work for other reasons) \cite{herrmann2010}.
 The presence of the fluctuations does not seem to be
related to the separation distance between the leads, as it was also found to take place,
 for instance, in the contact regime.
Similar oscillations have also been observed in NEGF calculations for carbon chains
(Fig.3 of  Ref.~\onlinecite{ozaki2010}), albeit with smaller amplitude. We also found that
replacing Pt with Au causes a reduction of the oscillations (see the corresponding
Mulliken population for the system $3-6-3$ in the inset of Fig.~\ref{fig:mulpo}). In this case,
the population did not seem to be affected by an improvement of the basis set as much as for Pt,
probably because of the different electronic structure.
Fig.~\ref{fig:mulpo} shows that, in order to keep a significant portion of the central region
 unaffected by such oscillations, it is necessary
to either increase $l$ ($r$)( as in panel d-e) or $c$ (as in panel b-c). However,
the second option is more recommendable as the first one
causes the computational time of the NEGF run to increase considerably.

\begin{figure}[t]
\begin{center}
\includegraphics[width=8.2cm]{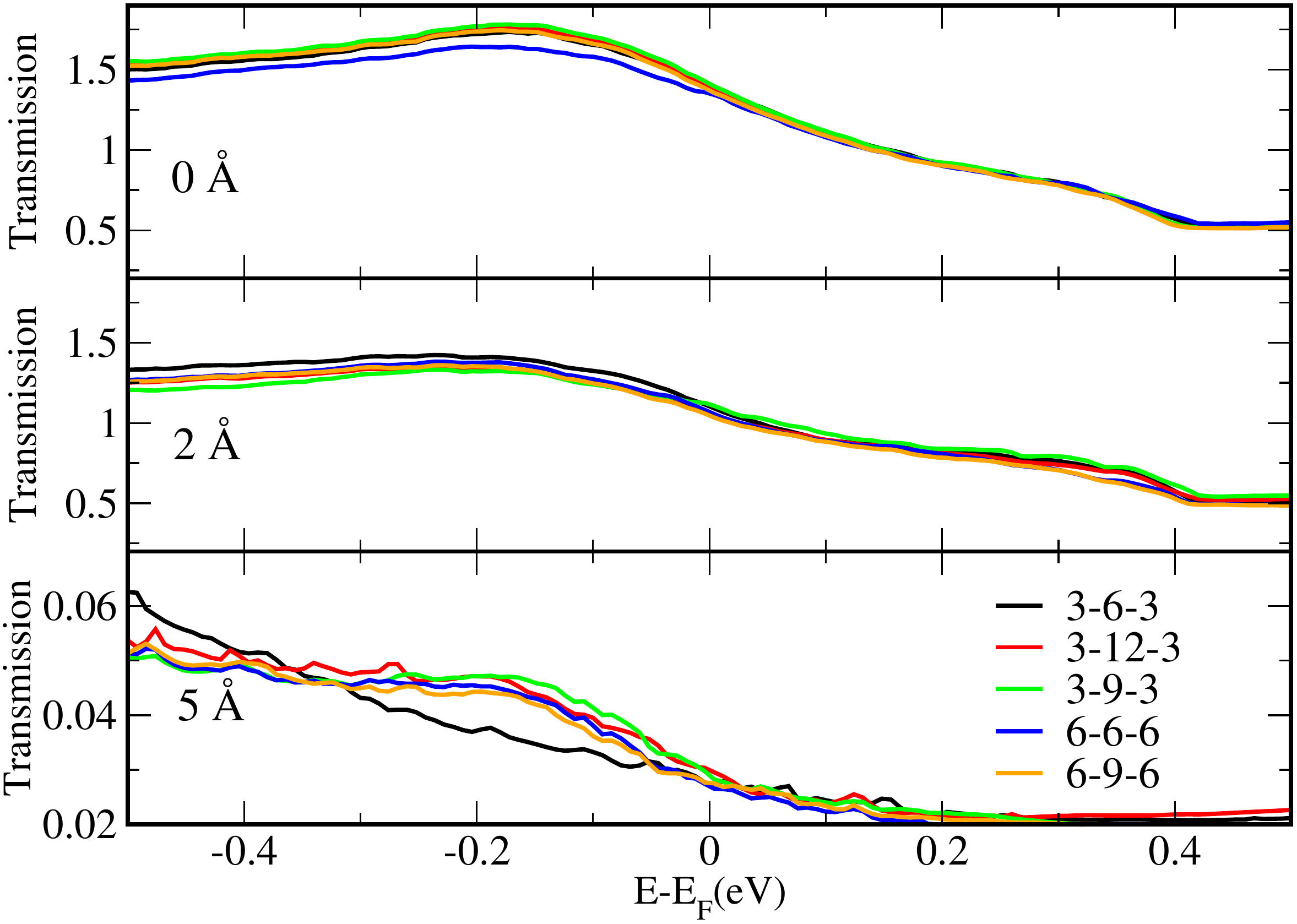}
\caption{(Color online) Transmission as a function of energy for the s2p2d1 systems (a-e) of
the right panels in Fig.~\ref{fig:mulpo}
} \label{fig:trasm-sl1x1}
\end{center}
\end{figure}

Subsequently, we proceeded to check how these charge oscillations affect the energy dependence
 of the electronic transmission. Fig.~\ref{fig:trasm-sl1x1} shows the transmission curves for
the systems (a)-(e) of the right column of Fig.~\ref{fig:mulpo}, calculated with the s2p2d1 basis set
 and  a 48$\times$48 $k$-point grid at three
different separation distances (0, 2 and 5~\AA). It can be observed how, at 0 and 2~\AA,
 no remarkable differences appear in the energy range around the Fermi level. At 5~\AA\
instead, in the same region the transmission
for the 3-6-3 system  is strikingly different from the others, as the transmission in the
energy region below is generally lower than in the curves of the  other $l-c-r$ sets.
Although the low$-$bias conductance values of the five transmission curves are not very different from
each other,  overall the results we presented so far indicate that  the
3-6-3 system is not large enough and either the size of the leads or of the central region must be
extended as in the other cases. Since extending the leads was found to  increase the computational
time more than increasing the central part did, we propose the  $3-12-3$ structure as a good
compromise.

\begin{figure}[t]
\begin{center}
\includegraphics[width=8.6cm]{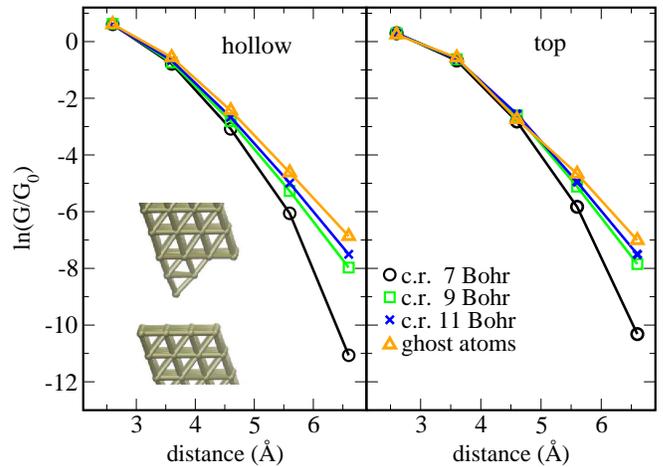}
\caption{$ln$ of the conductance obtained  throughout the stretching
process of the junctions, starting from a hollow and a top binding configuration,
 for three different cutoff radii as well as by making use of ghost atoms.
} \label{fig:decay}
\end{center}
\end{figure}

\begin{figure}[t]
\begin{center}
\includegraphics[width=8.6cm]{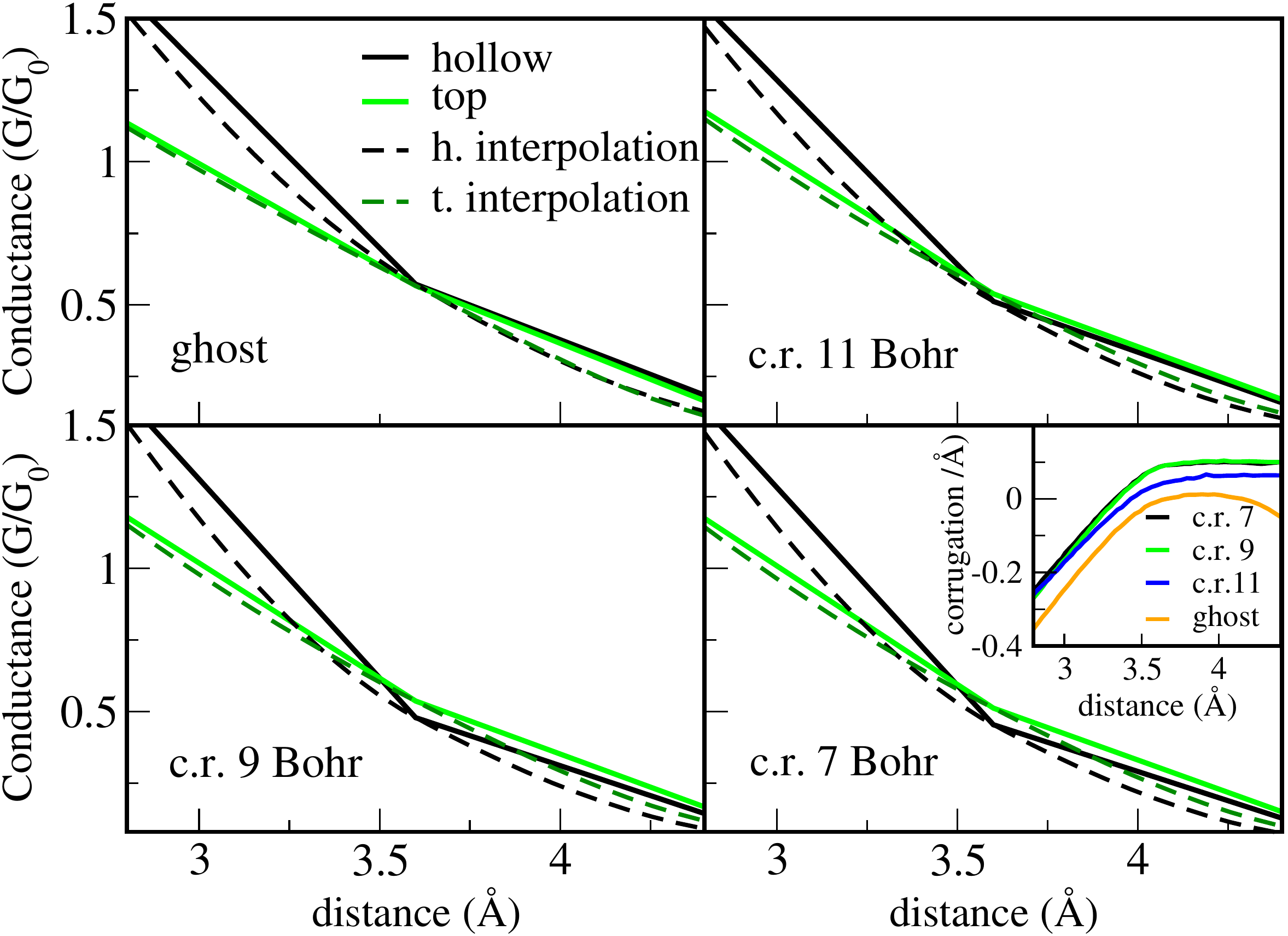}
\caption{Conductance in top and hollow geometry (solid lines) and corresponding interpolation
(dashed lines) for three different cutoff radii and for the insertion of ghost atoms. In the inset,
the corrugation as a function of the separation distance is shown.} \label{fig:interpolation}
\end{center}
\end{figure}

We now turn to analyze the distance dependence of the conductance and how it is affected by differences in the basis sets.
In the inset of  Fig.~\ref{fig:decay}, we show the geometry we employed. We considered two cases, in which the apex of the tip of the upper lead falls on an hcp and on a top site of the lower lead. The structure is a 4$\times$4 periodic cell and the central area comprises 6 layers on each side, while each lead consists of 3 layers.
In both cases, we increased  the separation distance from the initial value of 2.6~\AA\ and
 calculated the conductance stepwise. In Fig. ~\ref{fig:decay}, we compare the
 natural logarithm ($ln$) of the conductance values obtained in both cases.
We adopted three different cutoff radii for the basis set confinement (7, 9 and 11 Bohr).
 In the intermediate range 3.6 - 5.6 {\AA}, all curves approximately present a linear shape
 (corresponding to the expected  exponential decay of the conductance) and no significative
quantitative differences are visible. At closer distances we can observe the typical deviation
 from a linear behaviour, which is expected in the contact regime~\cite{garcia2010}.
For the shortest cutoff radius of 7 Bohr, the curves show deviation from linearity also
beyond 5.6 {\AA}. This problem is well known
and is usually solved by inserting ghost atoms in the vacuum region (which can occasionally
hamper the scf convergence) or by using
plane-wave-based methods~\cite{garcia2010,garcia2009optimal,GarciaLekue2015292}.
A comparison between the use of ghost atoms and increasing cutoff radius
was previously performed by Siesta calculations~\cite{garcia2009optimal} for what concerns the
 spatial behaviour of silver wave functions. There, the insertion of ghost atoms was claimed
 to provide closer results to those obtained with plane waves calculations. However, instead of
 the wave functions,  we chose to focus the comparison on a physical quantities such as
 the conductance and the corrugation. Notice also that, despite the
thorough study of Ref.~\onlinecite{garcia2009optimal}, the results there obtained cannot be
automatically applied to our system because of the differences in the
confinement scheme, as mentioned above.
In Fig.~\ref{fig:decay}, we also show the results obtained by inserting ghost atoms. It can be noticed that the ensuing
conductance values are very similar to those obtained by increasing the cutoff radii
to 11 Bohr. Moreover, the employment of ghost atoms did not show any particular advantage
concerning, for instance, the computational time needed to achieve convergence. This indicates that increasing the cutoff radius can indeed be
a viable alternative solution to this well-known problem, as it yields reliable results
without the insertion of additional elements in the geometry. We will also show in the
 following that the employment of ghost atoms can lead to a wrong evaluation of the corrugation  
in the tunneling regime. In Fig.~\ref{fig:interpolation}, we compare, for each cutoff radius and
 for the ghost-atom case, the conductance values obtained in the top geometry with that 
obtained in the  hollow at the same separation distance. Interpolation curves obtained by the Akima
 spline method in the range 2.6-4.6 {\AA} are also shown (dashed lines).
In the inset of the bottom right panel of the same figure we report the corrugation as a function
of distance for all four sets. This quantity was calculated from the interpolation curves by
 evaluating, at each conductance value, the difference in tip-surface distance between the
corresponding top and hollow geometries. Such a difference (the corrugation) has been plotted against 
the average  tip-surface distance. In all four cases, the corrugation is negative and follows a linear trend 
up to about a distance of 3.5 {\AA}. Beyond this value, however, while the curves for cutoff radii of 7, 9
 and 11 Bohr
reach positive values and saturate at 0.1 {\AA} approximately, the corrugation relative to the
 insertion of ghost atoms fails to reproduce this inversion, remaining negative.
The origin of such a discrepancy can be spotted in the main four panels of
 Fig.~\ref{fig:interpolation}: in the
ghost case, the conductance for the hollow position is always higher than for top,
whereas, in the other three cases, an inversion takes place at  3.5 \AA. This change 
at intermediate distances
is indeed expected, as it has also been observed in other systems \cite{kim2015site}.

\begin{figure}[t]
\begin{center}
\includegraphics[width=8.2cm]{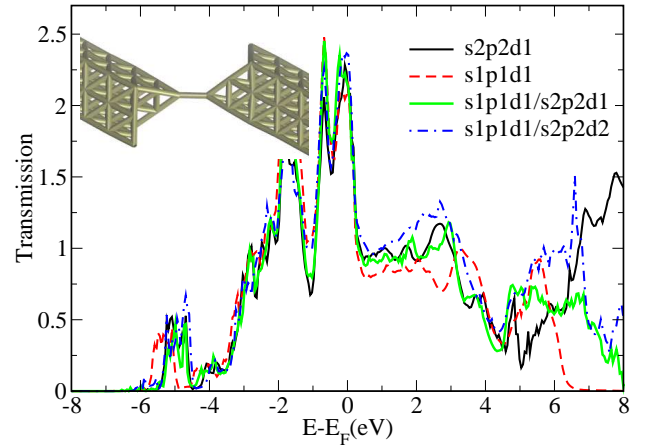}
\caption{(Color online) Transmission as a function of energy, for the geometry depicted
in the inset, in four different cases: in the first two,
either the s1p1d1 or the s2p2d1 basis set is used for both leads and central region;
 in the other two cases, the atoms in the tip and the innermost layer (on both sides)
is described by either s2p2d1 or s2p2d2, while the rest is treated by the s1p1d1 set.
} \label{fig:trasm-tiptip}
\end{center}
\end{figure}

Finally, in order to assess the reliability of our results, we explored how changes in the
number of primitive functions employed to construct the basis sets affect the conductance and how these results compare with those
obtained by other models.
To this aim, we chose the same geometry (depicted in the inset of Fig.~\ref{fig:trasm-tiptip}) as in Ref.~\onlinecite{thygesen2005},
where the Pt electronic structure was described by means of Wannier functions.
In our system, however, we added three more layers on each side of the central part
in order to avoid  the charge oscillations discussed above (Fig.~\ref{fig:mulpo}).
The distance between the two tip apices is 2.8~\AA, hence in the regime in which the
 conductance is not affected by the cutoff radius. Note that, for a similar structure,
 the lower coordination
of the tip apex was found to give rise to spin polarization.~\cite{Korytar201349}
However, for the present purpose we chose to neglect it since including it would worsen
the numerical performance considerably and would not change our main conclusions.
We either used the s1p1d1 or s2p2d1 basis set for all atoms. However, we also
 considered  a case in which not all atoms were treated in the same way: there, the atoms in the
 tip and the innermost layer on each side were described by either the s2p2d1 or s2p2d2 set,
 whereas the s1p1d1 set was used for the rest. This mixed strategy is quite frequent
in theoretical studies of metal nanocontacts as it allows a good description of
the central relevant region at a reasonable computational cost~\cite{wu2014,zotti2011}.
For all four cases, the self consistent cycle was performed with a 4$\times$4 $k$-point
grid; subsequently, the transmission, reported in Fig.~\ref{fig:trasm-tiptip}
as a function of energy, was calculated using the same 6$\times$6 $k$-point
 sampling as in Ref.~\onlinecite{thygesen2005}.
The four curves (see Fig.~\ref{fig:trasm-tiptip}) show, right below
the Fermi level, two peaks stemming from the expected contribution of  the $s$ and
 $d$ states~\cite{cuevas2003} and from the loss of degeneracy of the d states
due to symmetry reasons~\cite{nielsen2002}.
In the critical region around the Fermi level, the s1p1d1 curve shows a slight downshift (0.05 eV) with respect to
the other cases, in agreement with what was observed in Fig.~3 of Ref.~\onlinecite{strange2008}.
In this range  all curves
appear similar to those reported in Ref.~\onlinecite{thygesen2005}, although shifted
down in energy by around 0.25 eV.
In Appendix A, further comparisons with other models are shown. Overall, we can conclude that,
at the Fermi level and in the contact regime, detail differences in the basis sets do not affect either
 the general shape of
 the transmission curve or the energetic orbital alignment considerably. More noticeable changes were observed, instead,
at higher energies, which however do not affect the low bias conductance.   

\section{Conclusions} \label{sec-conclusions}
We studied the electronic transport through Pt nanocontacts in the $\left\{111\right\}$ orientation by using the combined
DFT+NEGF methods as implemented in OpenMX. We showed that, while the analysis of the contact regime
is not remarkably affected by details in the leads and in the construction of the basis sets, this
does not hold for the tunneling regime in many ways.  
For instance, at large separation distances
between the two electrodes forming the contact,
it is important to extend either the central or the lead region to a number of layers
so as to avoid the formation of unphysical charge oscillations in the region of interest.
We also showed that, in this regime, it is necessary to use basis sets with a large
cutoff radius to describe both the conductance decay and the corrugation accurately.
While the former could also be reproduced by employing ghost atoms,
the same strategy does not apply to the latter, for which ghost atoms fail to reproduce
the inversion in the corrugation expected at intermediate distances between top and hollow position.
These findings suggest that similar preliminary checks should be made for other materials,
whenever the performance of NEGF calculations aims at the study of distance-dependent effects.
This concerns, for example, the analysis of scanning-probe experiments.
Ultimately, we proved the robustness of our results by comparing them with those obtained
by other NEGF-based models.

\section{Acknowledgments}
We thank Eduardo Hernandez, Juan Carlos Cuevas and Juan Jose Palacios for fruitful
discussions and M. Todorovic for help with the DFT calculations. 
LAZ was funded by the Spanish MINECO through the grant MAT2014-58982-JIN.
RP acknowledges  financial support from MINECO projects CSD2010-00024, MAT2011-023627, 
MAT2014-54484-P, and MDM-2014-0377.
Computer time provided by the Spanish Supercomputing Network (RES) at the  MareNostrum III (BSC) and Tirant is
gratefully acknowledged.

\appendix

\begin{figure}[t]
\begin{center}
\includegraphics[width=8.2cm]{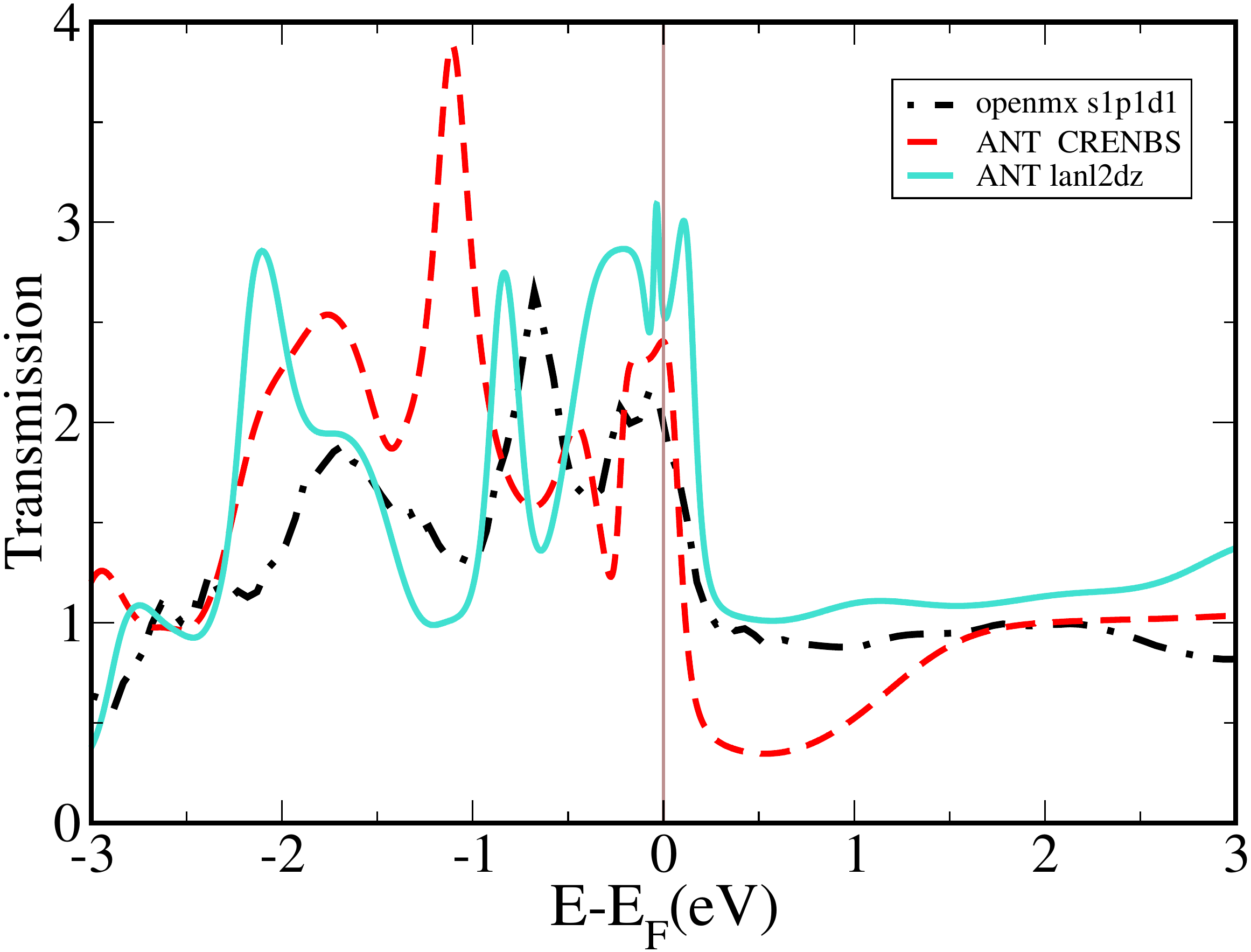}
\caption{(Color online) Comparison between the transmission curves obtained, for the
 geometry depicted in the inset of Fig.~\ref{fig:trasm-tiptip}, by OpenMX with the
 s1p1d1 basis set, and by ANT with the CRENBS and Lanl2dz basis set.
} \label{fig:comp-ant}
\end{center}
\end{figure}

\section{Comparison with ANT}
To further corroborate the robustness of our results, we investigated how the energy
 dependence of the transmission is affected by detail differences in the DFT+NEGF model employed.
 To this aim, we repeated our calculations by the quantum-chemistry code ANT
~\cite{Jacob2011,alacant}
 which is built as an interface to Gaussian~\cite{Gaussian}. There,
 metals are described by means of isolated clusters (thus no periodic boundary conditions
are applied as in OpenMX), while the basis sets consist of
linear combination of gaussian functions.
Combined DFT+NEGF methods are by now commonplace and have been previously
compared with other levels of theory~\cite{brandbyge2002}. However, to the best of our
 knowledge, comparisons between two codes which employ such fundamentally-different descriptions of
 the leads  are few and far between~\cite{pauly2008,bilan2012,strange2008}.
We chose to recompute the transmission for the structure of Fig.~\ref{fig:trasm-tiptip}. To this aim,
we employed  a geometry consisting of two  Au$_{20}$ pyramidal clusters.
 As for the basis set, we used a CRENBS~\cite{ross1990} basis set (which includes the same orbitals
 as in the s1p1d1 set of OpenMX) and a LANL2DZ set~\cite{wadt1985ab}.

\begin{figure}[hb]
\begin{center}
\includegraphics[width=8.2cm]{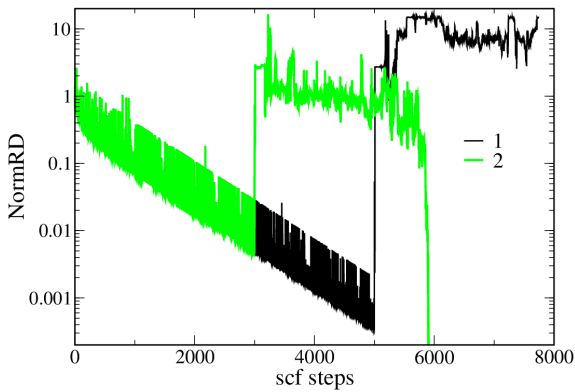}
\caption{(Color online) NormRD throughout the scf convergence steps for the case in which
the NEGF extension is switched on after 5000 steps (case 1, black curve) and  only 3000 steps (case 2, green curve).
Waiting for NormRD to reach a much lower value before switching to the NEGF method
does not accelerate the convergence process.
} \label{fig:per-paper-converg}
\end{center}
\end{figure}

In  Fig.~\ref{fig:comp-ant}, we show a comparison between the transmission
curves obtained by ANT and OpenMX.
 It can be observed that, in the range
 around the Fermi level, the shape of the OpenMX s1p1d1 curve and the ANT CRENBS curve
are quite similar, as well as the energy alignment. Differences arising
below this range are probably due to  how the core part is treated.
The LANl2dz basis set was previously claimed to give reliable results due to the fact
that it does not add ghost transmissions as supposidely-higher-quality basis sets do.
\cite{herrmann2010} Interestingly, the general shape of the curve obtained by OpenMX
with its minimal basis set appears to be quite similar to that calculated by ANT with
the  LANl2dz set. This indicates that, actually, the quality of the OpenMX minimal
basis set is comparable to that of well known higher-quality basis sets. However, the
 LANl2dz conductance values are higher by approximately 0.5 G/G$_{0}$.
The computed values can hardly be directly compared with the experiments, since
typical MCBJ measurements provide histograms which cannot be directly related
to exact detailed geometries. Experimental values around 0.5 and 1.5-2 G/G$_{0}$ have been reported
~\cite{evangeli2015,vardimon2014,nielsen2003} which do not necessary correspond to the
 geometry analyzed here. Indeed, because of the presence of the anisotropic $d$ states
at the Fermi level, the conductance is expected to be strongly influence by geometrical
 detail differences throughout the process
of formation and stretching of the junction. High-resolution transmission electron
microscopy indicate, for the geometry of Fig.~\ref{fig:trasm-tiptip}, a conductance
value of 4 G/G$_{0}$ (Fig.7e and Fig.8 of ref~\cite{kizuka2009}), which is closer to the LANl2dz
value.

\section{Scf convergence in OpenMX}
In the NEGF calculations as those discussed in this work, it is necessary to
achieve convergence between the input and output charge densities (let NormRD
be the residual norm between these two quantities). It is well known
that this is facilitated by performing a preliminary standard DFT run on the central region
to obtain the corresponding charge density as a starting point.
In  OpenMX, convergence can be achieved by four different mixing schemes.
Surprisingly, in all systems analyzed for this work,  the algorithm RMM-DIIS with Kerker's metric
\cite{rmdiisk}, which is slow but very efficient
 for standard DFT calculations, often failed to reduce NormRD in the NEGF runs.
Instead, convergence was eventually reached by using the Pulay algorithm~\cite{Pulay}.
In particular, we found it convenient to adopt a three-step procedure:
we first performed the preliminary  DFT calculation
by the RMM-DIIS algorithm until NormRD reached values smaller than at least
0.01; we then switched to using the Pulay algorithm for a few runs (about 10); we finally
switched to the NEGF technique by keeping the Pulay mixing scheme on.
Interestingly, we also observed that the number of steps necessary for convergence did not
always decrease by reducing NormRD to a much
 lower value than 0.01 (Fig. ~\ref{fig:per-paper-converg}).

\bibliography{literaturePtnanoc}

\begin{thebibliography}{63}%
\makeatletter
\providecommand \@ifxundefined [1]{%
 \@ifx{#1\undefined}
}%
\providecommand \@ifnum [1]{%
 \ifnum #1\expandafter \@firstoftwo
 \else \expandafter \@secondoftwo
 \fi
}%
\providecommand \@ifx [1]{%
 \ifx #1\expandafter \@firstoftwo
 \else \expandafter \@secondoftwo
 \fi
}%
\providecommand \natexlab [1]{#1}%
\providecommand \enquote  [1]{``#1''}%
\providecommand \bibnamefont  [1]{#1}%
\providecommand \bibfnamefont [1]{#1}%
\providecommand \citenamefont [1]{#1}%
\providecommand \href@noop [0]{\@secondoftwo}%
\providecommand \href [0]{\begingroup \@sanitize@url \@href}%
\providecommand \@href[1]{\@@startlink{#1}\@@href}%
\providecommand \@@href[1]{\endgroup#1\@@endlink}%
\providecommand \@sanitize@url [0]{\catcode `\\12\catcode `\$12\catcode
  `\&12\catcode `\#12\catcode `\^12\catcode `\_12\catcode `\%12\relax}%
\providecommand \@@startlink[1]{}%
\providecommand \@@endlink[0]{}%
\providecommand \url  [0]{\begingroup\@sanitize@url \@url }%
\providecommand \@url [1]{\endgroup\@href {#1}{\urlprefix }}%
\providecommand \urlprefix  [0]{URL }%
\providecommand \Eprint [0]{\href }%
\providecommand \doibase [0]{http://dx.doi.org/}%
\providecommand \selectlanguage [0]{\@gobble}%
\providecommand \bibinfo  [0]{\@secondoftwo}%
\providecommand \bibfield  [0]{\@secondoftwo}%
\providecommand \translation [1]{[#1]}%
\providecommand \BibitemOpen [0]{}%
\providecommand \bibitemStop [0]{}%
\providecommand \bibitemNoStop [0]{.\EOS\space}%
\providecommand \EOS [0]{\spacefactor3000\relax}%
\providecommand \BibitemShut  [1]{\csname bibitem#1\endcsname}%
\let\auto@bib@innerbib\@empty
\bibitem [{\citenamefont {Sattler}(2010)}]{sattler2010handbook}%
  \BibitemOpen
  \bibfield  {author} {\bibinfo {author} {\bibfnamefont {K.~D.}\ \bibnamefont
  {Sattler}},\ }\href@noop {} {\emph {\bibinfo {title} {Handbook of
  nanophysics: Principles and methods}}}\ (\bibinfo  {publisher} {CRC press},\
  \bibinfo {year} {2010})\BibitemShut {NoStop}%
\bibitem [{\citenamefont {Evangeli}\ \emph {et~al.}(2015)\citenamefont
  {Evangeli}, \citenamefont {Matt}, \citenamefont {Rinc{\'o}n-Garc{\'\i}a},
  \citenamefont {Pauly}, \citenamefont {Nielaba}, \citenamefont
  {Rubio-Bollinger}, \citenamefont {Cuevas},\ and\ \citenamefont
  {Agra{\"\i}t}}]{evangeli2015}%
  \BibitemOpen
  \bibfield  {author} {\bibinfo {author} {\bibfnamefont {C.}~\bibnamefont
  {Evangeli}}, \bibinfo {author} {\bibfnamefont {M.}~\bibnamefont {Matt}},
  \bibinfo {author} {\bibfnamefont {L.}~\bibnamefont {Rinc{\'o}n-Garc{\'\i}a}},
  \bibinfo {author} {\bibfnamefont {F.}~\bibnamefont {Pauly}}, \bibinfo
  {author} {\bibfnamefont {P.}~\bibnamefont {Nielaba}}, \bibinfo {author}
  {\bibfnamefont {G.}~\bibnamefont {Rubio-Bollinger}}, \bibinfo {author}
  {\bibfnamefont {J.~C.}\ \bibnamefont {Cuevas}}, \ and\ \bibinfo {author}
  {\bibfnamefont {N.}~\bibnamefont {Agra{\"\i}t}},\ }\href@noop {} {\bibfield
  {journal} {\bibinfo  {journal} {Nano Lett.}\ }\textbf {\bibinfo {volume}
  {15}},\ \bibinfo {pages} {1006} (\bibinfo {year} {2015})}\BibitemShut
  {NoStop}%
\bibitem [{\citenamefont {Cuevas}\ and\ \citenamefont
  {Scheer}(2010)}]{cuevas2010}%
  \BibitemOpen
  \bibfield  {author} {\bibinfo {author} {\bibfnamefont {J.~C.}\ \bibnamefont
  {Cuevas}}\ and\ \bibinfo {author} {\bibfnamefont {E.}~\bibnamefont
  {Scheer}},\ }\href@noop {} {\emph {\bibinfo {title} {Molecular electronics:
  an introduction to theory and experiment}}}\ (\bibinfo  {publisher} {World
  Scientific,Singapore},\ \bibinfo {year} {2010})\BibitemShut {NoStop}%
\bibitem [{\citenamefont {Jel\'{\i}nek}\ \emph {et~al.}(2005)\citenamefont
  {Jel\'{\i}nek}, \citenamefont {P{\'e}rez}, \citenamefont {Ortega},\ and\
  \citenamefont {Flores}}]{jelinek2005}%
  \BibitemOpen
  \bibfield  {author} {\bibinfo {author} {\bibfnamefont {P.}~\bibnamefont
  {Jel\'{\i}nek}}, \bibinfo {author} {\bibfnamefont {R.}~\bibnamefont
  {P{\'e}rez}}, \bibinfo {author} {\bibfnamefont {J.}~\bibnamefont {Ortega}}, \
  and\ \bibinfo {author} {\bibfnamefont {F.}~\bibnamefont {Flores}},\
  }\href@noop {} {\bibfield  {journal} {\bibinfo  {journal} {Nanotechnology}\
  }\textbf {\bibinfo {volume} {16}},\ \bibinfo {pages} {1023} (\bibinfo {year}
  {2005})}\BibitemShut {NoStop}%
\bibitem [{\citenamefont {Jel{\'\i}nek}\ \emph {et~al.}(2008)\citenamefont
  {Jel{\'\i}nek}, \citenamefont {P{\'e}rez}, \citenamefont {Ortega},\ and\
  \citenamefont {Flores}}]{jelinek2008}%
  \BibitemOpen
  \bibfield  {author} {\bibinfo {author} {\bibfnamefont {P.}~\bibnamefont
  {Jel{\'\i}nek}}, \bibinfo {author} {\bibfnamefont {R.}~\bibnamefont
  {P{\'e}rez}}, \bibinfo {author} {\bibfnamefont {J.}~\bibnamefont {Ortega}}, \
  and\ \bibinfo {author} {\bibfnamefont {F.}~\bibnamefont {Flores}},\
  }\href@noop {} {\bibfield  {journal} {\bibinfo  {journal} {Phys. Rev. B}\
  }\textbf {\bibinfo {volume} {77}},\ \bibinfo {pages} {115447} (\bibinfo
  {year} {2008})}\BibitemShut {NoStop}%
\bibitem [{\citenamefont {M\"uller}\ \emph {et~al.}(2016)\citenamefont
  {M\"uller}, \citenamefont {Salgado}, \citenamefont {N\'eel}, \citenamefont
  {Palacios},\ and\ \citenamefont {Kr\"oger}}]{Muller2016}%
  \BibitemOpen
  \bibfield  {author} {\bibinfo {author} {\bibfnamefont {M.}~\bibnamefont
  {M\"uller}}, \bibinfo {author} {\bibfnamefont {C.}~\bibnamefont {Salgado}},
  \bibinfo {author} {\bibfnamefont {N.}~\bibnamefont {N\'eel}}, \bibinfo
  {author} {\bibfnamefont {J.~J.}\ \bibnamefont {Palacios}}, \ and\ \bibinfo
  {author} {\bibfnamefont {J.}~\bibnamefont {Kr\"oger}},\ }\href {\doibase
  10.1103/PhysRevB.93.235402} {\bibfield  {journal} {\bibinfo  {journal} {Phys.
  Rev. B}\ }\textbf {\bibinfo {volume} {93}},\ \bibinfo {pages} {235402}
  (\bibinfo {year} {2016})}\BibitemShut {NoStop}%
\bibitem [{\citenamefont {Schirm}\ \emph {et~al.}(2013)\citenamefont {Schirm},
  \citenamefont {Matt}, \citenamefont {Pauly}, \citenamefont {Cuevas},
  \citenamefont {Nielaba},\ and\ \citenamefont {Scheer}}]{schirm2013}%
  \BibitemOpen
  \bibfield  {author} {\bibinfo {author} {\bibfnamefont {C.}~\bibnamefont
  {Schirm}}, \bibinfo {author} {\bibfnamefont {M.}~\bibnamefont {Matt}},
  \bibinfo {author} {\bibfnamefont {F.}~\bibnamefont {Pauly}}, \bibinfo
  {author} {\bibfnamefont {J.~C.}\ \bibnamefont {Cuevas}}, \bibinfo {author}
  {\bibfnamefont {P.}~\bibnamefont {Nielaba}}, \ and\ \bibinfo {author}
  {\bibfnamefont {E.}~\bibnamefont {Scheer}},\ }\href@noop {} {\bibfield
  {journal} {\bibinfo  {journal} {Nat. Nanotechnol.}\ }\textbf {\bibinfo
  {volume} {8}},\ \bibinfo {pages} {645} (\bibinfo {year} {2013})}\BibitemShut
  {NoStop}%
\bibitem [{\citenamefont {Fern{\'a}ndez-Rossier}\ \emph
  {et~al.}(2005)\citenamefont {Fern{\'a}ndez-Rossier}, \citenamefont {Jacob},
  \citenamefont {Untiedt},\ and\ \citenamefont {Palacios}}]{fernandez2005}%
  \BibitemOpen
  \bibfield  {author} {\bibinfo {author} {\bibfnamefont {J.}~\bibnamefont
  {Fern{\'a}ndez-Rossier}}, \bibinfo {author} {\bibfnamefont {D.}~\bibnamefont
  {Jacob}}, \bibinfo {author} {\bibfnamefont {C.}~\bibnamefont {Untiedt}}, \
  and\ \bibinfo {author} {\bibfnamefont {J.~J.}\ \bibnamefont {Palacios}},\
  }\href@noop {} {\bibfield  {journal} {\bibinfo  {journal} {Phys. Rev. B}\
  }\textbf {\bibinfo {volume} {72}},\ \bibinfo {pages} {224418} (\bibinfo
  {year} {2005})}\BibitemShut {NoStop}%
\bibitem [{\citenamefont {Garc\'{\i}a-Su{\'a}rez}\ \emph
  {et~al.}(2005)\citenamefont {Garc\'{\i}a-Su{\'a}rez}, \citenamefont {Rocha},
  \citenamefont {Bailey}, \citenamefont {Lambert}, \citenamefont {Sanvito},\
  and\ \citenamefont {Ferrer}}]{garcia2005}%
  \BibitemOpen
  \bibfield  {author} {\bibinfo {author} {\bibfnamefont {V.~M.}\ \bibnamefont
  {Garc\'{\i}a-Su{\'a}rez}}, \bibinfo {author} {\bibfnamefont {A.~R.}\
  \bibnamefont {Rocha}}, \bibinfo {author} {\bibfnamefont {S.~W.}\ \bibnamefont
  {Bailey}}, \bibinfo {author} {\bibfnamefont {C.~J.}\ \bibnamefont {Lambert}},
  \bibinfo {author} {\bibfnamefont {S.}~\bibnamefont {Sanvito}}, \ and\
  \bibinfo {author} {\bibfnamefont {J.}~\bibnamefont {Ferrer}},\ }\href@noop {}
  {\bibfield  {journal} {\bibinfo  {journal} {Phys. Rev. Lett.}\ }\textbf
  {\bibinfo {volume} {95}},\ \bibinfo {pages} {256804} (\bibinfo {year}
  {2005})}\BibitemShut {NoStop}%
\bibitem [{\citenamefont {Vardimon}\ \emph {et~al.}(2014)\citenamefont
  {Vardimon}, \citenamefont {Yelin}, \citenamefont {Klionsky}, \citenamefont
  {Sarkar}, \citenamefont {Biller}, \citenamefont {Kronik},\ and\ \citenamefont
  {Tal}}]{vardimon2014}%
  \BibitemOpen
  \bibfield  {author} {\bibinfo {author} {\bibfnamefont {R.}~\bibnamefont
  {Vardimon}}, \bibinfo {author} {\bibfnamefont {T.}~\bibnamefont {Yelin}},
  \bibinfo {author} {\bibfnamefont {M.}~\bibnamefont {Klionsky}}, \bibinfo
  {author} {\bibfnamefont {S.}~\bibnamefont {Sarkar}}, \bibinfo {author}
  {\bibfnamefont {A.~J.}\ \bibnamefont {Biller}}, \bibinfo {author}
  {\bibfnamefont {L.}~\bibnamefont {Kronik}}, \ and\ \bibinfo {author}
  {\bibfnamefont {O.}~\bibnamefont {Tal}},\ }\href@noop {} {\bibfield
  {journal} {\bibinfo  {journal} {Nano Lett.}\ }\textbf {\bibinfo {volume}
  {14}},\ \bibinfo {pages} {2988} (\bibinfo {year} {2014})}\BibitemShut
  {NoStop}%
\bibitem [{\citenamefont {Parashar}\ \emph {et~al.}(2014)\citenamefont
  {Parashar}, \citenamefont {Srivastava}, \citenamefont {Pattanaik},\ and\
  \citenamefont {Jain}}]{parashar2014}%
  \BibitemOpen
  \bibfield  {author} {\bibinfo {author} {\bibfnamefont {S.}~\bibnamefont
  {Parashar}}, \bibinfo {author} {\bibfnamefont {P.}~\bibnamefont
  {Srivastava}}, \bibinfo {author} {\bibfnamefont {M.}~\bibnamefont
  {Pattanaik}}, \ and\ \bibinfo {author} {\bibfnamefont {S.~K.}\ \bibnamefont
  {Jain}},\ }\href@noop {} {\bibfield  {journal} {\bibinfo  {journal} {Eur.
  Phys. J. B}\ }\textbf {\bibinfo {volume} {87}},\ \bibinfo {pages} {1}
  (\bibinfo {year} {2014})}\BibitemShut {NoStop}%
\bibitem [{\citenamefont {Makk}\ \emph {et~al.}(2012)\citenamefont {Makk},
  \citenamefont {Balogh}, \citenamefont {Csonka},\ and\ \citenamefont
  {Halbritter}}]{makk2012pulling}%
  \BibitemOpen
  \bibfield  {author} {\bibinfo {author} {\bibfnamefont {P.}~\bibnamefont
  {Makk}}, \bibinfo {author} {\bibfnamefont {Z.}~\bibnamefont {Balogh}},
  \bibinfo {author} {\bibfnamefont {S.}~\bibnamefont {Csonka}}, \ and\ \bibinfo
  {author} {\bibfnamefont {A.}~\bibnamefont {Halbritter}},\ }\href@noop {}
  {\bibfield  {journal} {\bibinfo  {journal} {Nanoscale}\ }\textbf {\bibinfo
  {volume} {4}},\ \bibinfo {pages} {4739} (\bibinfo {year} {2012})}\BibitemShut
  {NoStop}%
\bibitem [{\citenamefont {Garc\'{\i}a}\ \emph {et~al.}(2004)\citenamefont
  {Garc\'{\i}a}, \citenamefont {Palacios}, \citenamefont {SanFabi\'an},
  \citenamefont {Verg\'es}, \citenamefont {P\'erez-Jim\'enez},\ and\
  \citenamefont {Louis}}]{garcia2004}%
  \BibitemOpen
  \bibfield  {author} {\bibinfo {author} {\bibfnamefont {Y.}~\bibnamefont
  {Garc\'{\i}a}}, \bibinfo {author} {\bibfnamefont {J.~J.}\ \bibnamefont
  {Palacios}}, \bibinfo {author} {\bibfnamefont {E.}~\bibnamefont
  {SanFabi\'an}}, \bibinfo {author} {\bibfnamefont {J.~A.}\ \bibnamefont
  {Verg\'es}}, \bibinfo {author} {\bibfnamefont {A.~J.}\ \bibnamefont
  {P\'erez-Jim\'enez}}, \ and\ \bibinfo {author} {\bibfnamefont
  {E.}~\bibnamefont {Louis}},\ }\href@noop {} {\bibfield  {journal} {\bibinfo
  {journal} {Phys. Rev. B}\ }\textbf {\bibinfo {volume} {69}},\ \bibinfo
  {pages} {041402} (\bibinfo {year} {2004})}\BibitemShut {NoStop}%
\bibitem [{\citenamefont {Strange}\ \emph {et~al.}(2006)\citenamefont
  {Strange}, \citenamefont {Thygesen},\ and\ \citenamefont
  {Jacobsen}}]{strange2006}%
  \BibitemOpen
  \bibfield  {author} {\bibinfo {author} {\bibfnamefont {M.}~\bibnamefont
  {Strange}}, \bibinfo {author} {\bibfnamefont {K.~S.}\ \bibnamefont
  {Thygesen}}, \ and\ \bibinfo {author} {\bibfnamefont {K.~W.}\ \bibnamefont
  {Jacobsen}},\ }\href@noop {} {\bibfield  {journal} {\bibinfo  {journal}
  {Phys. Rev. B}\ }\textbf {\bibinfo {volume} {73}},\ \bibinfo {pages} {125424}
  (\bibinfo {year} {2006})}\BibitemShut {NoStop}%
\bibitem [{\citenamefont {Strange}\ \emph {et~al.}(2008)\citenamefont
  {Strange}, \citenamefont {Kristensen}, \citenamefont {Thygesen},\ and\
  \citenamefont {Jacobsen}}]{strange2008}%
  \BibitemOpen
  \bibfield  {author} {\bibinfo {author} {\bibfnamefont {M.}~\bibnamefont
  {Strange}}, \bibinfo {author} {\bibfnamefont {I.}~\bibnamefont {Kristensen}},
  \bibinfo {author} {\bibfnamefont {K.~S.}\ \bibnamefont {Thygesen}}, \ and\
  \bibinfo {author} {\bibfnamefont {K.~W.}\ \bibnamefont {Jacobsen}},\
  }\href@noop {} {\bibfield  {journal} {\bibinfo  {journal} {J. Chem. Phys.}\
  }\textbf {\bibinfo {volume} {128}},\ \bibinfo {pages} {114714} (\bibinfo
  {year} {2008})}\BibitemShut {NoStop}%
\bibitem [{\citenamefont {Cuevas}\ \emph {et~al.}(2003)\citenamefont {Cuevas},
  \citenamefont {Heurich}, \citenamefont {Pauly}, \citenamefont {Wenzel},\ and\
  \citenamefont {Sch{\"o}n}}]{cuevas2003}%
  \BibitemOpen
  \bibfield  {author} {\bibinfo {author} {\bibfnamefont {J.~C.}\ \bibnamefont
  {Cuevas}}, \bibinfo {author} {\bibfnamefont {J.}~\bibnamefont {Heurich}},
  \bibinfo {author} {\bibfnamefont {F.}~\bibnamefont {Pauly}}, \bibinfo
  {author} {\bibfnamefont {W.}~\bibnamefont {Wenzel}}, \ and\ \bibinfo {author}
  {\bibfnamefont {G.}~\bibnamefont {Sch{\"o}n}},\ }\href@noop {} {\bibfield
  {journal} {\bibinfo  {journal} {Nanotechnology}\ }\textbf {\bibinfo {volume}
  {14}},\ \bibinfo {pages} {R29} (\bibinfo {year} {2003})}\BibitemShut
  {NoStop}%
\bibitem [{\citenamefont {Zhang}\ \emph {et~al.}(2010)\citenamefont {Zhang},
  \citenamefont {Ma}, \citenamefont {Bai}, \citenamefont {Sun}, \citenamefont
  {Rungger}, \citenamefont {Shen}, \citenamefont {Sanvito},\ and\ \citenamefont
  {Hou}}]{zhang2010}%
  \BibitemOpen
  \bibfield  {author} {\bibinfo {author} {\bibfnamefont {R.}~\bibnamefont
  {Zhang}}, \bibinfo {author} {\bibfnamefont {G.}~\bibnamefont {Ma}}, \bibinfo
  {author} {\bibfnamefont {M.}~\bibnamefont {Bai}}, \bibinfo {author}
  {\bibfnamefont {L.}~\bibnamefont {Sun}}, \bibinfo {author} {\bibfnamefont
  {I.}~\bibnamefont {Rungger}}, \bibinfo {author} {\bibfnamefont
  {Z.}~\bibnamefont {Shen}}, \bibinfo {author} {\bibfnamefont {S.}~\bibnamefont
  {Sanvito}}, \ and\ \bibinfo {author} {\bibfnamefont {S.}~\bibnamefont
  {Hou}},\ }\href@noop {} {\bibfield  {journal} {\bibinfo  {journal}
  {Nanotechnology}\ }\textbf {\bibinfo {volume} {21}},\ \bibinfo {pages}
  {155203} (\bibinfo {year} {2010})}\BibitemShut {NoStop}%
\bibitem [{\citenamefont {Nielsen}\ \emph {et~al.}(2003)\citenamefont
  {Nielsen}, \citenamefont {Noat}, \citenamefont {Brandbyge}, \citenamefont
  {Smit}, \citenamefont {Hansen}, \citenamefont {Chen}, \citenamefont {Yanson},
  \citenamefont {Besenbacher},\ and\ \citenamefont {van
  Ruitenbeek}}]{nielsen2003}%
  \BibitemOpen
  \bibfield  {author} {\bibinfo {author} {\bibfnamefont {S.~K.}\ \bibnamefont
  {Nielsen}}, \bibinfo {author} {\bibfnamefont {Y.}~\bibnamefont {Noat}},
  \bibinfo {author} {\bibfnamefont {M.}~\bibnamefont {Brandbyge}}, \bibinfo
  {author} {\bibfnamefont {R.~H.~M.}\ \bibnamefont {Smit}}, \bibinfo {author}
  {\bibfnamefont {K.}~\bibnamefont {Hansen}}, \bibinfo {author} {\bibfnamefont
  {L.~Y.}\ \bibnamefont {Chen}}, \bibinfo {author} {\bibfnamefont {A.~I.}\
  \bibnamefont {Yanson}}, \bibinfo {author} {\bibfnamefont {F.}~\bibnamefont
  {Besenbacher}}, \ and\ \bibinfo {author} {\bibfnamefont {J.~M.}\ \bibnamefont
  {van Ruitenbeek}},\ }\href@noop {} {\bibfield  {journal} {\bibinfo  {journal}
  {Phys. Rev. B}\ }\textbf {\bibinfo {volume} {67}},\ \bibinfo {pages} {245411}
  (\bibinfo {year} {2003})}\BibitemShut {NoStop}%
\bibitem [{\citenamefont {Nielsen}\ \emph {et~al.}(2002)\citenamefont
  {Nielsen}, \citenamefont {Brandbyge}, \citenamefont {Hansen}, \citenamefont
  {Stokbro}, \citenamefont {van Ruitenbeek},\ and\ \citenamefont
  {Besenbacher}}]{nielsen2002}%
  \BibitemOpen
  \bibfield  {author} {\bibinfo {author} {\bibfnamefont {S.~K.}\ \bibnamefont
  {Nielsen}}, \bibinfo {author} {\bibfnamefont {M.}~\bibnamefont {Brandbyge}},
  \bibinfo {author} {\bibfnamefont {K.}~\bibnamefont {Hansen}}, \bibinfo
  {author} {\bibfnamefont {K.}~\bibnamefont {Stokbro}}, \bibinfo {author}
  {\bibfnamefont {J.~M.}\ \bibnamefont {van Ruitenbeek}}, \ and\ \bibinfo
  {author} {\bibfnamefont {F.}~\bibnamefont {Besenbacher}},\ }\href@noop {}
  {\bibfield  {journal} {\bibinfo  {journal} {Phys. Rev. Lett.}\ }\textbf
  {\bibinfo {volume} {89}},\ \bibinfo {pages} {066804} (\bibinfo {year}
  {2002})}\BibitemShut {NoStop}%
\bibitem [{\citenamefont {Wu}\ \emph {et~al.}(2014)\citenamefont {Wu},
  \citenamefont {Bai}, \citenamefont {Sanvito},\ and\ \citenamefont
  {Hou}}]{wu2014}%
  \BibitemOpen
  \bibfield  {author} {\bibinfo {author} {\bibfnamefont {K.}~\bibnamefont
  {Wu}}, \bibinfo {author} {\bibfnamefont {M.}~\bibnamefont {Bai}}, \bibinfo
  {author} {\bibfnamefont {S.}~\bibnamefont {Sanvito}}, \ and\ \bibinfo
  {author} {\bibfnamefont {S.}~\bibnamefont {Hou}},\ }\href@noop {} {\bibfield
  {journal} {\bibinfo  {journal} {J. Chem. Phys.}\ }\textbf {\bibinfo {volume}
  {141}},\ \bibinfo {pages} {014707} (\bibinfo {year} {2014})}\BibitemShut
  {NoStop}%
\bibitem [{\citenamefont {Thygesen}\ and\ \citenamefont
  {Jacobsen}(2005)}]{thygesen2005}%
  \BibitemOpen
  \bibfield  {author} {\bibinfo {author} {\bibfnamefont {K.~S.}\ \bibnamefont
  {Thygesen}}\ and\ \bibinfo {author} {\bibfnamefont {K.~W.}\ \bibnamefont
  {Jacobsen}},\ }\href@noop {} {\bibfield  {journal} {\bibinfo  {journal}
  {Phys. Rev. B}\ }\textbf {\bibinfo {volume} {72}},\ \bibinfo {pages} {033401}
  (\bibinfo {year} {2005})}\BibitemShut {NoStop}%
\bibitem [{\citenamefont {Pauly}\ \emph {et~al.}(2006)\citenamefont {Pauly},
  \citenamefont {Dreher}, \citenamefont {Viljas}, \citenamefont {H{\"a}fner},
  \citenamefont {Cuevas},\ and\ \citenamefont {Nielaba}}]{pauly2006}%
  \BibitemOpen
  \bibfield  {author} {\bibinfo {author} {\bibfnamefont {F.}~\bibnamefont
  {Pauly}}, \bibinfo {author} {\bibfnamefont {M.}~\bibnamefont {Dreher}},
  \bibinfo {author} {\bibfnamefont {J.~K.}\ \bibnamefont {Viljas}}, \bibinfo
  {author} {\bibfnamefont {M.}~\bibnamefont {H{\"a}fner}}, \bibinfo {author}
  {\bibfnamefont {J.~C.}\ \bibnamefont {Cuevas}}, \ and\ \bibinfo {author}
  {\bibfnamefont {P.}~\bibnamefont {Nielaba}},\ }\href@noop {} {\bibfield
  {journal} {\bibinfo  {journal} {Phys. Rev. B}\ }\textbf {\bibinfo {volume}
  {74}},\ \bibinfo {pages} {235106} (\bibinfo {year} {2006})}\BibitemShut
  {NoStop}%
\bibitem [{\citenamefont {Smogunov}\ \emph {et~al.}(2008)\citenamefont
  {Smogunov}, \citenamefont {Dal~Corso},\ and\ \citenamefont
  {Tosatti}}]{smogunov2008}%
  \BibitemOpen
  \bibfield  {author} {\bibinfo {author} {\bibfnamefont {A.}~\bibnamefont
  {Smogunov}}, \bibinfo {author} {\bibfnamefont {A.}~\bibnamefont {Dal~Corso}},
  \ and\ \bibinfo {author} {\bibfnamefont {E.}~\bibnamefont {Tosatti}},\
  }\href@noop {} {\bibfield  {journal} {\bibinfo  {journal} {Phys. Rev. B}\
  }\textbf {\bibinfo {volume} {78}},\ \bibinfo {pages} {014423} (\bibinfo
  {year} {2008})}\BibitemShut {NoStop}%
\bibitem [{\citenamefont {Ono}(2009)}]{Ono2009}%
  \BibitemOpen
  \bibfield  {author} {\bibinfo {author} {\bibfnamefont {T.}~\bibnamefont
  {Ono}},\ }\href@noop {} {\bibfield  {journal} {\bibinfo  {journal} {Jour.
  Phys. Chem. C}\ }\textbf {\bibinfo {volume} {113}},\ \bibinfo {pages} {6256}
  (\bibinfo {year} {2009})}\BibitemShut {NoStop}%
\bibitem [{\citenamefont {Pauly~F}\ and\ \citenamefont {G}(2008)}]{Pauly2008a}%
  \BibitemOpen
  \bibfield  {author} {\bibinfo {author} {\bibfnamefont {H.~U. H. M. W. S. B.
  M. C. J.~C.}\ \bibnamefont {Pauly~F}, \bibfnamefont {Viljas J~K}}\ and\
  \bibinfo {author} {\bibfnamefont {S.}~\bibnamefont {G}},\ }\href@noop {}
  {\bibfield  {journal} {\bibinfo  {journal} {New. J. Phys.}\ }\textbf
  {\bibinfo {volume} {10}},\ \bibinfo {pages} {125019} (\bibinfo {year}
  {2008})}\BibitemShut {NoStop}%
\bibitem [{\citenamefont {Palacios}\ \emph {et~al.}()\citenamefont {Palacios},
  \citenamefont {Jacob}, \citenamefont {Perez-Jimenez}, \citenamefont {Fabian},
  \citenamefont {Louis},\ and\ \citenamefont {Verges}}]{alacant}%
  \BibitemOpen
  \bibfield  {author} {\bibinfo {author} {\bibfnamefont {J.~J.}\ \bibnamefont
  {Palacios}}, \bibinfo {author} {\bibfnamefont {D.}~\bibnamefont {Jacob}},
  \bibinfo {author} {\bibfnamefont {A.~J.}\ \bibnamefont {Perez-Jimenez}},
  \bibinfo {author} {\bibfnamefont {E.~S.}\ \bibnamefont {Fabian}}, \bibinfo
  {author} {\bibfnamefont {E.}~\bibnamefont {Louis}}, \ and\ \bibinfo {author}
  {\bibfnamefont {J.~A.}\ \bibnamefont {Verges}},\ }\href@noop {} {\bibinfo
  {journal} {ALACANT ab initio quantum transport package, see URL:
  http://alacant.dfa.ua.es}\ }\BibitemShut {NoStop}%
\bibitem [{\citenamefont {Brandbyge}\ \emph {et~al.}(2002)\citenamefont
  {Brandbyge}, \citenamefont {Mozos}, \citenamefont {Ordej{\'o}n},
  \citenamefont {Taylor},\ and\ \citenamefont {Stokbro}}]{brandbyge2002}%
  \BibitemOpen
\bibfield  {journal} {  }\bibfield  {author} {\bibinfo {author} {\bibfnamefont
  {M.}~\bibnamefont {Brandbyge}}, \bibinfo {author} {\bibfnamefont {J.-L.}\
  \bibnamefont {Mozos}}, \bibinfo {author} {\bibfnamefont {P.}~\bibnamefont
  {Ordej{\'o}n}}, \bibinfo {author} {\bibfnamefont {J.}~\bibnamefont {Taylor}},
  \ and\ \bibinfo {author} {\bibfnamefont {K.}~\bibnamefont {Stokbro}},\
  }\href@noop {} {\bibfield  {journal} {\bibinfo  {journal} {Phys. Rev. B}\
  }\textbf {\bibinfo {volume} {65}},\ \bibinfo {pages} {165401} (\bibinfo
  {year} {2002})}\BibitemShut {NoStop}%
\bibitem [{\citenamefont {Rocha}\ \emph {et~al.}(2006)\citenamefont {Rocha},
  \citenamefont {Garc\'{\i}a-Su{\'a}rez}, \citenamefont {Bailey}, \citenamefont
  {Lambert}, \citenamefont {Ferrer},\ and\ \citenamefont
  {Sanvito}}]{rocha2006spin}%
  \BibitemOpen
  \bibfield  {author} {\bibinfo {author} {\bibfnamefont {A.~R.}\ \bibnamefont
  {Rocha}}, \bibinfo {author} {\bibfnamefont {V.~M.}\ \bibnamefont
  {Garc\'{\i}a-Su{\'a}rez}}, \bibinfo {author} {\bibfnamefont {S.}~\bibnamefont
  {Bailey}}, \bibinfo {author} {\bibfnamefont {C.}~\bibnamefont {Lambert}},
  \bibinfo {author} {\bibfnamefont {J.}~\bibnamefont {Ferrer}}, \ and\ \bibinfo
  {author} {\bibfnamefont {S.}~\bibnamefont {Sanvito}},\ }\href@noop {}
  {\bibfield  {journal} {\bibinfo  {journal} {Phys. Rev. B}\ }\textbf {\bibinfo
  {volume} {73}},\ \bibinfo {pages} {085414} (\bibinfo {year}
  {2006})}\BibitemShut {NoStop}%
\bibitem [{\citenamefont {Ozaki}\ \emph {et~al.}(2010)\citenamefont {Ozaki},
  \citenamefont {Nishio},\ and\ \citenamefont {Kino}}]{ozaki2010}%
  \BibitemOpen
  \bibfield  {author} {\bibinfo {author} {\bibfnamefont {T.}~\bibnamefont
  {Ozaki}}, \bibinfo {author} {\bibfnamefont {K.}~\bibnamefont {Nishio}}, \
  and\ \bibinfo {author} {\bibfnamefont {H.}~\bibnamefont {Kino}},\ }\href@noop
  {} {\bibfield  {journal} {\bibinfo  {journal} {Phys. Rev. B}\ }\textbf
  {\bibinfo {volume} {81}},\ \bibinfo {pages} {035116} (\bibinfo {year}
  {2010})}\BibitemShut {NoStop}%
\bibitem [{\citenamefont {Xue}\ \emph {et~al.}(2001)\citenamefont {Xue},
  \citenamefont {Datta},\ and\ \citenamefont {Ratner}}]{Xue2001}%
  \BibitemOpen
  \bibfield  {author} {\bibinfo {author} {\bibfnamefont {Y.}~\bibnamefont
  {Xue}}, \bibinfo {author} {\bibfnamefont {S.}~\bibnamefont {Datta}}, \ and\
  \bibinfo {author} {\bibfnamefont {M.~A.}\ \bibnamefont {Ratner}},\
  }\href@noop {} {\bibfield  {journal} {\bibinfo  {journal} {J. Chem. Phys.}\
  }\textbf {\bibinfo {volume} {115}} (\bibinfo {year} {2001})}\BibitemShut
  {NoStop}%
\bibitem [{\citenamefont {Ferrer}\ \emph {et~al.}(2014)\citenamefont {Ferrer},
  \citenamefont {Lambert}, \citenamefont {Garc{\'\i}a-Su{\'a}rez},
  \citenamefont {Manrique}, \citenamefont {Visontai}, \citenamefont
  {Oroszlany}, \citenamefont {Rodr{\'\i}guez-Ferrad{\'a}s}, \citenamefont
  {Grace}, \citenamefont {Bailey}, \citenamefont {Gillemot} \emph
  {et~al.}}]{ferrer2014}%
  \BibitemOpen
  \bibfield  {author} {\bibinfo {author} {\bibfnamefont {J.}~\bibnamefont
  {Ferrer}}, \bibinfo {author} {\bibfnamefont {C.~J.}\ \bibnamefont {Lambert}},
  \bibinfo {author} {\bibfnamefont {V.~M.}\ \bibnamefont
  {Garc{\'\i}a-Su{\'a}rez}}, \bibinfo {author} {\bibfnamefont {D.~Z.}\
  \bibnamefont {Manrique}}, \bibinfo {author} {\bibfnamefont {D.}~\bibnamefont
  {Visontai}}, \bibinfo {author} {\bibfnamefont {L.}~\bibnamefont {Oroszlany}},
  \bibinfo {author} {\bibfnamefont {R.}~\bibnamefont
  {Rodr{\'\i}guez-Ferrad{\'a}s}}, \bibinfo {author} {\bibfnamefont
  {I.}~\bibnamefont {Grace}}, \bibinfo {author} {\bibfnamefont
  {S.}~\bibnamefont {Bailey}}, \bibinfo {author} {\bibfnamefont
  {K.}~\bibnamefont {Gillemot}},  \emph {et~al.},\ }\href@noop {} {\bibfield
  {journal} {\bibinfo  {journal} {New. J. Phys.}\ }\textbf {\bibinfo {volume}
  {16}},\ \bibinfo {pages} {093029} (\bibinfo {year} {2014})}\BibitemShut
  {NoStop}%
\bibitem [{\citenamefont {Garcia-Lekue}\ \emph {et~al.}(2015)\citenamefont
  {Garcia-Lekue}, \citenamefont {Vergniory}, \citenamefont {Jiang},\ and\
  \citenamefont {Wang}}]{GarciaLekue2015292}%
  \BibitemOpen
  \bibfield  {author} {\bibinfo {author} {\bibfnamefont {A.}~\bibnamefont
  {Garcia-Lekue}}, \bibinfo {author} {\bibfnamefont {M.}~\bibnamefont
  {Vergniory}}, \bibinfo {author} {\bibfnamefont {X.}~\bibnamefont {Jiang}}, \
  and\ \bibinfo {author} {\bibfnamefont {L.}~\bibnamefont {Wang}},\ }\href
  {\doibase http://dx.doi.org/10.1016/j.progsurf.2015.05.002} {\bibfield
  {journal} {\bibinfo  {journal} {Progress in Surface Science}\ }\textbf
  {\bibinfo {volume} {90}},\ \bibinfo {pages} {292 } (\bibinfo {year}
  {2015})}\BibitemShut {NoStop}%
\bibitem [{\citenamefont {Quek}\ \emph {et~al.}(2007)\citenamefont {Quek},
  \citenamefont {Venkataraman}, \citenamefont {Choi}, \citenamefont {Louie},
  \citenamefont {Hybertsen},\ and\ \citenamefont {B}}]{Quek2007}%
  \BibitemOpen
  \bibfield  {author} {\bibinfo {author} {\bibfnamefont {S.~Y.}\ \bibnamefont
  {Quek}}, \bibinfo {author} {\bibfnamefont {L.}~\bibnamefont {Venkataraman}},
  \bibinfo {author} {\bibfnamefont {H.~J.}\ \bibnamefont {Choi}}, \bibinfo
  {author} {\bibfnamefont {S.~G.}\ \bibnamefont {Louie}}, \bibinfo {author}
  {\bibfnamefont {M.~S.}\ \bibnamefont {Hybertsen}}, \ and\ \bibinfo {author}
  {\bibfnamefont {N.~J.}\ \bibnamefont {B}},\ }\href@noop {} {\bibfield
  {journal} {\bibinfo  {journal} {Nano Lett.}\ }\textbf {\bibinfo {volume}
  {7}},\ \bibinfo {pages} {3477} (\bibinfo {year} {2007})}\BibitemShut
  {NoStop}%
\bibitem [{\citenamefont {Limot}\ \emph {et~al.}(2005)\citenamefont {Limot},
  \citenamefont {Kr{\"o}ger}, \citenamefont {Berndt}, \citenamefont
  {Garcia-Lekue},\ and\ \citenamefont {Hofer}}]{limot2005}%
  \BibitemOpen
  \bibfield  {author} {\bibinfo {author} {\bibfnamefont {L.}~\bibnamefont
  {Limot}}, \bibinfo {author} {\bibfnamefont {J.}~\bibnamefont {Kr{\"o}ger}},
  \bibinfo {author} {\bibfnamefont {R.}~\bibnamefont {Berndt}}, \bibinfo
  {author} {\bibfnamefont {A.}~\bibnamefont {Garcia-Lekue}}, \ and\ \bibinfo
  {author} {\bibfnamefont {W.~A.}\ \bibnamefont {Hofer}},\ }\href@noop {}
  {\bibfield  {journal} {\bibinfo  {journal} {Phys. Rev. Lett.}\ }\textbf
  {\bibinfo {volume} {94}},\ \bibinfo {pages} {126102} (\bibinfo {year}
  {2005})}\BibitemShut {NoStop}%
\bibitem [{\citenamefont {Garcia-Lekue}\ and\ \citenamefont
  {Wang}(2010)}]{garcia2010}%
  \BibitemOpen
  \bibfield  {author} {\bibinfo {author} {\bibfnamefont {A.}~\bibnamefont
  {Garcia-Lekue}}\ and\ \bibinfo {author} {\bibfnamefont {L.~W.}\ \bibnamefont
  {Wang}},\ }\href@noop {} {\bibfield  {journal} {\bibinfo  {journal} {Phys.
  Rev. B}\ }\textbf {\bibinfo {volume} {82}},\ \bibinfo {pages} {035410}
  (\bibinfo {year} {2010})}\BibitemShut {NoStop}%
\bibitem [{\citenamefont {Frederiksen}\ \emph {et~al.}(2007)\citenamefont
  {Frederiksen}, \citenamefont {Lorente}, \citenamefont {Paulsson},\ and\
  \citenamefont {Brandbyge}}]{Frederiksen2007}%
  \BibitemOpen
  \bibfield  {author} {\bibinfo {author} {\bibfnamefont {T.}~\bibnamefont
  {Frederiksen}}, \bibinfo {author} {\bibfnamefont {N.}~\bibnamefont
  {Lorente}}, \bibinfo {author} {\bibfnamefont {M.}~\bibnamefont {Paulsson}}, \
  and\ \bibinfo {author} {\bibfnamefont {M.}~\bibnamefont {Brandbyge}},\ }\href
  {\doibase 10.1103/PhysRevB.75.235441} {\bibfield  {journal} {\bibinfo
  {journal} {Phys. Rev. B}\ }\textbf {\bibinfo {volume} {75}},\ \bibinfo
  {pages} {235441} (\bibinfo {year} {2007})}\BibitemShut {NoStop}%
\bibitem [{\citenamefont {Polok}\ \emph {et~al.}(2011)\citenamefont {Polok},
  \citenamefont {Fedorov}, \citenamefont {Bagrets}, \citenamefont {Zahn},\ and\
  \citenamefont {Mertig}}]{Polok2011}%
  \BibitemOpen
  \bibfield  {author} {\bibinfo {author} {\bibfnamefont {M.}~\bibnamefont
  {Polok}}, \bibinfo {author} {\bibfnamefont {D.~V.}\ \bibnamefont {Fedorov}},
  \bibinfo {author} {\bibfnamefont {A.}~\bibnamefont {Bagrets}}, \bibinfo
  {author} {\bibfnamefont {P.}~\bibnamefont {Zahn}}, \ and\ \bibinfo {author}
  {\bibfnamefont {I.}~\bibnamefont {Mertig}},\ }\href {\doibase
  10.1103/PhysRevB.83.245426} {\bibfield  {journal} {\bibinfo  {journal} {Phys.
  Rev. B}\ }\textbf {\bibinfo {volume} {83}},\ \bibinfo {pages} {245426}
  (\bibinfo {year} {2011})}\BibitemShut {NoStop}%
\bibitem [{\citenamefont {Sharma}\ \emph {et~al.}(2013)\citenamefont {Sharma},
  \citenamefont {Ansari}, \citenamefont {Feldman}, \citenamefont {Iakovidis},
  \citenamefont {Greer},\ and\ \citenamefont {Fagas}}]{sharma2013}%
  \BibitemOpen
  \bibfield  {author} {\bibinfo {author} {\bibfnamefont {D.}~\bibnamefont
  {Sharma}}, \bibinfo {author} {\bibfnamefont {L.}~\bibnamefont {Ansari}},
  \bibinfo {author} {\bibfnamefont {B.}~\bibnamefont {Feldman}}, \bibinfo
  {author} {\bibfnamefont {M.}~\bibnamefont {Iakovidis}}, \bibinfo {author}
  {\bibfnamefont {J.}~\bibnamefont {Greer}}, \ and\ \bibinfo {author}
  {\bibfnamefont {G.}~\bibnamefont {Fagas}},\ }\href@noop {} {\bibfield
  {journal} {\bibinfo  {journal} {J. Appl. Phys.}\ }\textbf {\bibinfo {volume}
  {113}},\ \bibinfo {pages} {203708} (\bibinfo {year} {2013})}\BibitemShut
  {NoStop}%
\bibitem [{\citenamefont {Okuno}\ and\ \citenamefont
  {Ozaki}(2012)}]{okuno2012}%
  \BibitemOpen
  \bibfield  {author} {\bibinfo {author} {\bibfnamefont {Y.}~\bibnamefont
  {Okuno}}\ and\ \bibinfo {author} {\bibfnamefont {T.}~\bibnamefont {Ozaki}},\
  }\href@noop {} {\bibfield  {journal} {\bibinfo  {journal} {Jour. Phys. Chem.
  C}\ }\textbf {\bibinfo {volume} {117}},\ \bibinfo {pages} {100} (\bibinfo
  {year} {2012})}\BibitemShut {NoStop}%
\bibitem [{\citenamefont {Ansari}\ \emph {et~al.}(2012)\citenamefont {Ansari},
  \citenamefont {Fagas}, \citenamefont {Colinge},\ and\ \citenamefont
  {Greer}}]{ansari2012}%
  \BibitemOpen
  \bibfield  {author} {\bibinfo {author} {\bibfnamefont {L.}~\bibnamefont
  {Ansari}}, \bibinfo {author} {\bibfnamefont {G.}~\bibnamefont {Fagas}},
  \bibinfo {author} {\bibfnamefont {J.-P.}\ \bibnamefont {Colinge}}, \ and\
  \bibinfo {author} {\bibfnamefont {J.~C.}\ \bibnamefont {Greer}},\ }\href@noop
  {} {\bibfield  {journal} {\bibinfo  {journal} {Nano letters}\ }\textbf
  {\bibinfo {volume} {12}},\ \bibinfo {pages} {2222} (\bibinfo {year}
  {2012})}\BibitemShut {NoStop}%
\bibitem [{\citenamefont {Lan}\ \emph {et~al.}(2015)\citenamefont {Lan},
  \citenamefont {Ho},\ and\ \citenamefont {Hai}}]{lan2015electronic}%
  \BibitemOpen
  \bibfield  {author} {\bibinfo {author} {\bibfnamefont {T.~N.}\ \bibnamefont
  {Lan}}, \bibinfo {author} {\bibfnamefont {L.~B.}\ \bibnamefont {Ho}}, \ and\
  \bibinfo {author} {\bibfnamefont {T.~H.}\ \bibnamefont {Hai}},\ }\href@noop
  {} {\bibfield  {journal} {\bibinfo  {journal} {physica status solidi (b)}\
  }\textbf {\bibinfo {volume} {252}},\ \bibinfo {pages} {573} (\bibinfo {year}
  {2015})}\BibitemShut {NoStop}%
\bibitem [{\citenamefont {Hashmi}\ \emph {et~al.}(2015)\citenamefont {Hashmi},
  \citenamefont {Farooq}, \citenamefont {Hu},\ and\ \citenamefont
  {Hong}}]{hashmi2015spin}%
  \BibitemOpen
  \bibfield  {author} {\bibinfo {author} {\bibfnamefont {A.}~\bibnamefont
  {Hashmi}}, \bibinfo {author} {\bibfnamefont {M.~U.}\ \bibnamefont {Farooq}},
  \bibinfo {author} {\bibfnamefont {T.}~\bibnamefont {Hu}}, \ and\ \bibinfo
  {author} {\bibfnamefont {J.}~\bibnamefont {Hong}},\ }\href@noop {} {\bibfield
   {journal} {\bibinfo  {journal} {Jour. Phys. Chem. C}\ }\textbf {\bibinfo
  {volume} {119}},\ \bibinfo {pages} {1859} (\bibinfo {year}
  {2015})}\BibitemShut {NoStop}%
\bibitem [{\citenamefont {Jippo}\ \emph {et~al.}(2015)\citenamefont {Jippo},
  \citenamefont {Ohfuchi},\ and\ \citenamefont {Okada}}]{jippo2015electronic}%
  \BibitemOpen
  \bibfield  {author} {\bibinfo {author} {\bibfnamefont {H.}~\bibnamefont
  {Jippo}}, \bibinfo {author} {\bibfnamefont {M.}~\bibnamefont {Ohfuchi}}, \
  and\ \bibinfo {author} {\bibfnamefont {S.}~\bibnamefont {Okada}},\
  }\href@noop {} {\bibfield  {journal} {\bibinfo  {journal} {e-J. Surf. Sci.
  Nanotech.}\ }\textbf {\bibinfo {volume} {13}},\ \bibinfo {pages} {54}
  (\bibinfo {year} {2015})}\BibitemShut {NoStop}%
\bibitem [{\citenamefont {Ozaki}(2003)}]{ozaki2003}%
  \BibitemOpen
  \bibfield  {author} {\bibinfo {author} {\bibfnamefont {T.}~\bibnamefont
  {Ozaki}},\ }\href@noop {} {\bibfield  {journal} {\bibinfo  {journal} {Phys.
  Rev. B}\ }\textbf {\bibinfo {volume} {67}},\ \bibinfo {pages} {155108}
  (\bibinfo {year} {2003})}\BibitemShut {NoStop}%
\bibitem [{\citenamefont {Ozaki}\ and\ \citenamefont {Kino}(2004)}]{ozaki2004}%
  \BibitemOpen
  \bibfield  {author} {\bibinfo {author} {\bibfnamefont {T.}~\bibnamefont
  {Ozaki}}\ and\ \bibinfo {author} {\bibfnamefont {H.}~\bibnamefont {Kino}},\
  }\href@noop {} {\bibfield  {journal} {\bibinfo  {journal} {Phys. Rev. B}\
  }\textbf {\bibinfo {volume} {69}},\ \bibinfo {pages} {195113} (\bibinfo
  {year} {2004})}\BibitemShut {NoStop}%
\bibitem [{\citenamefont {Perdew}\ \emph {et~al.}(1997)\citenamefont {Perdew},
  \citenamefont {Burke},\ and\ \citenamefont {Ernzerhof}}]{perdew1997}%
  \BibitemOpen
  \bibfield  {author} {\bibinfo {author} {\bibfnamefont {J.~P.}\ \bibnamefont
  {Perdew}}, \bibinfo {author} {\bibfnamefont {K.}~\bibnamefont {Burke}}, \
  and\ \bibinfo {author} {\bibfnamefont {M.}~\bibnamefont {Ernzerhof}},\
  }\href@noop {} {\bibfield  {journal} {\bibinfo  {journal} {Phys. Rev. Lett.}\
  }\textbf {\bibinfo {volume} {78}},\ \bibinfo {pages} {1396} (\bibinfo {year}
  {1997})}\BibitemShut {NoStop}%
\bibitem [{\citenamefont {Monkhorst}\ and\ \citenamefont
  {Pack}(1976)}]{Monkhorst}%
  \BibitemOpen
  \bibfield  {author} {\bibinfo {author} {\bibfnamefont {H.~J.}\ \bibnamefont
  {Monkhorst}}\ and\ \bibinfo {author} {\bibfnamefont {J.~D.}\ \bibnamefont
  {Pack}},\ }\href@noop {} {\bibfield  {journal} {\bibinfo  {journal} {Phys.
  Rev. B}\ }\textbf {\bibinfo {volume} {13}},\ \bibinfo {pages} {5188}
  (\bibinfo {year} {1976})}\BibitemShut {NoStop}%
\bibitem [{ATK()}]{ATK}%
  \BibitemOpen
  \href@noop {} {\bibinfo  {journal} {Atomistix ToolKit, QuantumWise A/S
  (2012). http://quantumwise.com/documents/manuals}\ }\BibitemShut {NoStop}%
\bibitem [{\citenamefont {Junquera}\ \emph {et~al.}(2001)\citenamefont
  {Junquera}, \citenamefont {Paz}, \citenamefont {S{\'a}nchez-Portal},\ and\
  \citenamefont {Artacho}}]{junquera2001}%
  \BibitemOpen
\bibfield  {journal} {  }\bibfield  {author} {\bibinfo {author} {\bibfnamefont
  {J.}~\bibnamefont {Junquera}}, \bibinfo {author} {\bibfnamefont
  {{\'O}.}~\bibnamefont {Paz}}, \bibinfo {author} {\bibfnamefont
  {D.}~\bibnamefont {S{\'a}nchez-Portal}}, \ and\ \bibinfo {author}
  {\bibfnamefont {E.}~\bibnamefont {Artacho}},\ }\href@noop {} {\bibfield
  {journal} {\bibinfo  {journal} {Phys. Rev. B}\ }\textbf {\bibinfo {volume}
  {64}},\ \bibinfo {pages} {235111} (\bibinfo {year} {2001})}\BibitemShut
  {NoStop}%
\bibitem [{\citenamefont {Herrmann}\ \emph {et~al.}(2010)\citenamefont
  {Herrmann}, \citenamefont {Solomon}, \citenamefont {Subotnik}, \citenamefont
  {Mujica},\ and\ \citenamefont {Ratner}}]{herrmann2010}%
  \BibitemOpen
  \bibfield  {author} {\bibinfo {author} {\bibfnamefont {C.}~\bibnamefont
  {Herrmann}}, \bibinfo {author} {\bibfnamefont {G.~C.}\ \bibnamefont
  {Solomon}}, \bibinfo {author} {\bibfnamefont {J.~E.}\ \bibnamefont
  {Subotnik}}, \bibinfo {author} {\bibfnamefont {V.}~\bibnamefont {Mujica}}, \
  and\ \bibinfo {author} {\bibfnamefont {M.~A.}\ \bibnamefont {Ratner}},\
  }\href@noop {} {\bibfield  {journal} {\bibinfo  {journal} {J. Chem. Phys.}\
  }\textbf {\bibinfo {volume} {132}},\ \bibinfo {pages} {024103} (\bibinfo
  {year} {2010})}\BibitemShut {NoStop}%
\bibitem [{\citenamefont {Garc\'{\i}a-Gil}\ \emph {et~al.}(2009)\citenamefont
  {Garc\'{\i}a-Gil}, \citenamefont {Garc\'{\i}a}, \citenamefont {Lorente},\
  and\ \citenamefont {Ordejon}}]{garcia2009optimal}%
  \BibitemOpen
  \bibfield  {author} {\bibinfo {author} {\bibfnamefont {S.}~\bibnamefont
  {Garc\'{\i}a-Gil}}, \bibinfo {author} {\bibfnamefont {A.}~\bibnamefont
  {Garc\'{\i}a}}, \bibinfo {author} {\bibfnamefont {N.}~\bibnamefont
  {Lorente}}, \ and\ \bibinfo {author} {\bibfnamefont {P.}~\bibnamefont
  {Ordejon}},\ }\href@noop {} {\bibfield  {journal} {\bibinfo  {journal} {Phys.
  Rev. B}\ }\textbf {\bibinfo {volume} {79}},\ \bibinfo {pages} {075441}
  (\bibinfo {year} {2009})}\BibitemShut {NoStop}%
\bibitem [{\citenamefont {Kim}\ and\ \citenamefont
  {Hasegawa}(2015)}]{kim2015site}%
  \BibitemOpen
  \bibfield  {author} {\bibinfo {author} {\bibfnamefont {H.}~\bibnamefont
  {Kim}}\ and\ \bibinfo {author} {\bibfnamefont {Y.}~\bibnamefont {Hasegawa}},\
  }\href@noop {} {\bibfield  {journal} {\bibinfo  {journal} {Phys. Rev. Lett.}\
  }\textbf {\bibinfo {volume} {114}},\ \bibinfo {pages} {206801} (\bibinfo
  {year} {2015})}\BibitemShut {NoStop}%
\bibitem [{\citenamefont {Korytár}\ and\ \citenamefont
  {Evers}(2013)}]{Korytar201349}%
  \BibitemOpen
  \bibfield  {author} {\bibinfo {author} {\bibfnamefont {R.}~\bibnamefont
  {Korytár}}\ and\ \bibinfo {author} {\bibfnamefont {F.}~\bibnamefont
  {Evers}},\ }\href@noop {} {\bibfield  {journal} {\bibinfo  {journal} {Surface
  Science}\ }\textbf {\bibinfo {volume} {618}},\ \bibinfo {pages} {49}
  (\bibinfo {year} {2013})}\BibitemShut {NoStop}%
\bibitem [{\citenamefont {Zotti}\ \emph {et~al.}(2011)\citenamefont {Zotti},
  \citenamefont {B{\"u}rkle}, \citenamefont {Dappe}, \citenamefont {Pauly},\
  and\ \citenamefont {Cuevas}}]{zotti2011}%
  \BibitemOpen
  \bibfield  {author} {\bibinfo {author} {\bibfnamefont {L.~A.}\ \bibnamefont
  {Zotti}}, \bibinfo {author} {\bibfnamefont {M.}~\bibnamefont {B{\"u}rkle}},
  \bibinfo {author} {\bibfnamefont {Y.~J.}\ \bibnamefont {Dappe}}, \bibinfo
  {author} {\bibfnamefont {F.}~\bibnamefont {Pauly}}, \ and\ \bibinfo {author}
  {\bibfnamefont {J.~C.}\ \bibnamefont {Cuevas}},\ }\href@noop {} {\bibfield
  {journal} {\bibinfo  {journal} {Phys. Rev. B}\ }\textbf {\bibinfo {volume}
  {84}},\ \bibinfo {pages} {193404} (\bibinfo {year} {2011})}\BibitemShut
  {NoStop}%
\bibitem [{\citenamefont {Jacob}\ and\ \citenamefont
  {Palacios}(2011)}]{Jacob2011}%
  \BibitemOpen
  \bibfield  {author} {\bibinfo {author} {\bibfnamefont {D.}~\bibnamefont
  {Jacob}}\ and\ \bibinfo {author} {\bibfnamefont {J.~J.}\ \bibnamefont
  {Palacios}},\ }\href@noop {} {\bibfield  {journal} {\bibinfo  {journal} {J.
  Chem. Phys.}\ }\textbf {\bibinfo {volume} {134}},\ \bibinfo {pages} {044118}
  (\bibinfo {year} {2011})}\BibitemShut {NoStop}%
\bibitem [{\citenamefont {Frish}\ \emph {et~al.}(2003)\citenamefont {Frish},
  \citenamefont {Trucks},\ and\ \citenamefont {et~al.}}]{Gaussian}%
  \BibitemOpen
  \bibfield  {author} {\bibinfo {author} {\bibfnamefont {M.~J.}\ \bibnamefont
  {Frish}}, \bibinfo {author} {\bibfnamefont {G.~W.}\ \bibnamefont {Trucks}}, \
  and\ \bibinfo {author} {\bibfnamefont {H.~B.~S.}\ \bibnamefont {et~al.}},\
  }\href@noop {} {\bibfield  {journal} {\bibinfo  {journal} {GAUSSIAN 03,
  Revision B.01,Gaussian. Inc., Pittsburg}\ } (\bibinfo {year}
  {2003})}\BibitemShut {NoStop}%
\bibitem [{\citenamefont {Pauly}\ \emph {et~al.}(2008)\citenamefont {Pauly},
  \citenamefont {Viljas}, \citenamefont {Huniar}, \citenamefont {H{\"a}fner},
  \citenamefont {Wohlthat}, \citenamefont {B{\"u}rkle}, \citenamefont
  {Cuevas},\ and\ \citenamefont {Sch{\"o}n}}]{pauly2008}%
  \BibitemOpen
  \bibfield  {author} {\bibinfo {author} {\bibfnamefont {F.}~\bibnamefont
  {Pauly}}, \bibinfo {author} {\bibfnamefont {J.~K.}\ \bibnamefont {Viljas}},
  \bibinfo {author} {\bibfnamefont {U.}~\bibnamefont {Huniar}}, \bibinfo
  {author} {\bibfnamefont {M.}~\bibnamefont {H{\"a}fner}}, \bibinfo {author}
  {\bibfnamefont {S.}~\bibnamefont {Wohlthat}}, \bibinfo {author}
  {\bibfnamefont {M.}~\bibnamefont {B{\"u}rkle}}, \bibinfo {author}
  {\bibfnamefont {J.~C.}\ \bibnamefont {Cuevas}}, \ and\ \bibinfo {author}
  {\bibfnamefont {G.}~\bibnamefont {Sch{\"o}n}},\ }\href@noop {} {\bibfield
  {journal} {\bibinfo  {journal} {New. J. Phys.}\ }\textbf {\bibinfo {volume}
  {10}},\ \bibinfo {pages} {125019} (\bibinfo {year} {2008})}\BibitemShut
  {NoStop}%
\bibitem [{\citenamefont {Bilan}\ \emph {et~al.}(2012)\citenamefont {Bilan},
  \citenamefont {Zotti}, \citenamefont {Pauly},\ and\ \citenamefont
  {Cuevas}}]{bilan2012}%
  \BibitemOpen
  \bibfield  {author} {\bibinfo {author} {\bibfnamefont {S.}~\bibnamefont
  {Bilan}}, \bibinfo {author} {\bibfnamefont {L.~A.}\ \bibnamefont {Zotti}},
  \bibinfo {author} {\bibfnamefont {F.}~\bibnamefont {Pauly}}, \ and\ \bibinfo
  {author} {\bibfnamefont {J.~C.}\ \bibnamefont {Cuevas}},\ }\href@noop {}
  {\bibfield  {journal} {\bibinfo  {journal} {Phys. Rev. B}\ }\textbf {\bibinfo
  {volume} {85}},\ \bibinfo {pages} {205403} (\bibinfo {year}
  {2012})}\BibitemShut {NoStop}%
\bibitem [{\citenamefont {Ross}\ \emph {et~al.}(1990)\citenamefont {Ross},
  \citenamefont {Powers}, \citenamefont {Atashroo}, \citenamefont {Ermler},
  \citenamefont {LaJohn},\ and\ \citenamefont {Christiansen}}]{ross1990}%
  \BibitemOpen
  \bibfield  {author} {\bibinfo {author} {\bibfnamefont {R.}~\bibnamefont
  {Ross}}, \bibinfo {author} {\bibfnamefont {J.}~\bibnamefont {Powers}},
  \bibinfo {author} {\bibfnamefont {T.}~\bibnamefont {Atashroo}}, \bibinfo
  {author} {\bibfnamefont {W.}~\bibnamefont {Ermler}}, \bibinfo {author}
  {\bibfnamefont {L.}~\bibnamefont {LaJohn}}, \ and\ \bibinfo {author}
  {\bibfnamefont {P.}~\bibnamefont {Christiansen}},\ }\href@noop {} {\bibfield
  {journal} {\bibinfo  {journal} {J. Chem. Phys.}\ }\textbf {\bibinfo {volume}
  {93}},\ \bibinfo {pages} {6654} (\bibinfo {year} {1990})}\BibitemShut
  {NoStop}%
\bibitem [{\citenamefont {Wadt}\ and\ \citenamefont {Hay}(1985)}]{wadt1985ab}%
  \BibitemOpen
  \bibfield  {author} {\bibinfo {author} {\bibfnamefont {W.~R.}\ \bibnamefont
  {Wadt}}\ and\ \bibinfo {author} {\bibfnamefont {P.~J.}\ \bibnamefont {Hay}},\
  }\href@noop {} {\bibfield  {journal} {\bibinfo  {journal} {J. Chem. Phys.}\
  }\textbf {\bibinfo {volume} {82}},\ \bibinfo {pages} {284} (\bibinfo {year}
  {1985})}\BibitemShut {NoStop}%
\bibitem [{\citenamefont {Kizuka}\ and\ \citenamefont
  {Monna}(2009)}]{kizuka2009}%
  \BibitemOpen
  \bibfield  {author} {\bibinfo {author} {\bibfnamefont {T.}~\bibnamefont
  {Kizuka}}\ and\ \bibinfo {author} {\bibfnamefont {K.}~\bibnamefont {Monna}},\
  }\href@noop {} {\bibfield  {journal} {\bibinfo  {journal} {Phys. Rev. B}\
  }\textbf {\bibinfo {volume} {80}},\ \bibinfo {pages} {205406} (\bibinfo
  {year} {2009})}\BibitemShut {NoStop}%
\bibitem [{\citenamefont {Kresse}\ and\ \citenamefont
  {Furthmuller}(1996)}]{rmdiisk}%
  \BibitemOpen
  \bibfield  {author} {\bibinfo {author} {\bibfnamefont {G.}~\bibnamefont
  {Kresse}}\ and\ \bibinfo {author} {\bibfnamefont {J.}~\bibnamefont
  {Furthmuller}},\ }\href@noop {} {\bibfield  {journal} {\bibinfo  {journal}
  {Phys. Rev. B}\ }\textbf {\bibinfo {volume} {54}},\ \bibinfo {pages} {11169}
  (\bibinfo {year} {1996})}\BibitemShut {NoStop}%
\bibitem [{\citenamefont {Bowler}\ and\ \citenamefont {Gillan}(2000)}]{Pulay}%
  \BibitemOpen
  \bibfield  {author} {\bibinfo {author} {\bibfnamefont {D.~R.}\ \bibnamefont
  {Bowler}}\ and\ \bibinfo {author} {\bibfnamefont {M.~J.}\ \bibnamefont
  {Gillan}},\ }\href@noop {} {\bibfield  {journal} {\bibinfo  {journal} {Chem.
  Phys. Lett.}\ }\textbf {\bibinfo {volume} {325}},\ \bibinfo {pages} {475}
  (\bibinfo {year} {2000})}\BibitemShut {NoStop}%
\end{thebibliography}%

\end{document}